\documentclass[11pt,a4paper,oneside]{amsart}
\pdfoutput=1

\usepackage[a4paper,textwidth=15.5cm,textheight=23cm,centering]{geometry}
\usepackage{cite}

\usepackage[]{amsmath}
	\numberwithin{equation}{section}
\usepackage[]{amssymb}
\usepackage{amsthm}
	\theoremstyle{remark}
	\newtheorem*{remark}{Remark}
\usepackage{amsfonts}

\usepackage{enumerate}

\newcommand{\Id}{1\!\!1}
\newcommand{\I}{\mathrm{i}}

\DeclareMathOperator{\R}{\mathbb{R}}

\DeclareMathOperator{\N}{\mathbb{N}}
\DeclareMathOperator{\Z}{\mathbb{Z}}
\DeclareMathOperator{\cS}{\mathbb{S}}
\DeclareMathOperator{\tr}{\mathrm{Tr}}
\DeclareMathOperator{\la2}{\mathfrak{su}\left(2\right)}
\DeclareMathOperator{\Str}{\mathrm{STr}}

\newcommand{\mz}{\mathcal{Z}}

\usepackage[utf8]{inputenc}
\usepackage[english]{babel}

\usepackage[]{hyperref}

\usepackage[pdftex]{graphicx}
\usepackage{xcolor}

\usepackage[toc,page]{appendix}

\begin{document}
\title[]{Complex (super)-matrix models with external sources and $q$-ensembles of Chern--Simons and ABJ(M) type}
\author{Leonardo Santilli}
\address[LS]{Grupo de F\'{i}sica Matem\'{a}tica, Departamento de Matem\'{a}tica, Faculdade de Ci\^{e}ncias, Universidade de Lisboa, Campo Grande, Edif\'{i}cio C6, 1749-016 Lisboa, Portugal.}
\email{lsantilli@fc.ul.pt}

\author{Miguel Tierz}
\address[MT]{Departamento de Matem\'{a}tica, Faculdade de Ci\^{e}ncias, ISCTE - Instituto Universit\'{a}rio de Lisboa, Avenida das For\c{c}as Armadas, 1649-026 Lisboa, Portugal.}
\email{mtpaz@iscte-iul.pt}
\address[MT]{Grupo de F\'{i}sica Matem\'{a}tica, Departamento de Matem\'{a}tica, Faculdade de Ci\^{e}ncias, Universidade de Lisboa, Campo Grande, Edif\'{i}cio C6, 1749-016 Lisboa, Portugal.}
\email{tierz@fc.ul.pt}

\begin{abstract}
The Langmann--Szabo--Zarembo (LSZ) matrix model is a complex matrix model
with a quartic interaction and two external matrices. The model appears in
the study of a scalar field theory on the non-commutative plane. We
prove that the LSZ matrix model computes the probability of atypically large
fluctuations in the Stieltjes--Wigert matrix model, which is a $q$-ensemble
describing $U(N)$ Chern--Simons theory on the three-sphere. The correspondence holds in a
generalized sense: depending on the spectra of the two external matrices,
the LSZ matrix model either describes probabilities of large fluctuations in
the Chern--Simons partition function, in the unknot invariant or in the
two-unknot invariant. We extend the result to supermatrix models, and show
that a generalized LSZ supermatrix model describes the probability of
atypically large fluctuations in the ABJ(M) matrix model.
\end{abstract}

\maketitle

\tableofcontents

\section{Introduction}

The idea of introducing non-zero commutators among position or momentum
coordinates in different directions goes back to the 1940s \cite{Snyder:1947}. 
The consequences of applying ideas and results of noncommutative geometry
in quantum field theory are far-reaching but also considerably involved with
both several non-trivial results and difficulties as well.\par

In the late 1990s there was a boost of interest in noncommutative (NC) field
theories in great part due to the fact that low energy string theory can be
related to NC field theory \cite{SeibergWitten:1999,LizziSzabo:1999}.
However, soon it was established that the expectation that washing out the
space-time points could weaken UV divergences in quantum field theory, and
consequently simplify renormalization, did not work as expected. Rather the
opposite turned out to hold: renormalization gets harder due to
noncommutativity, because in planar diagrams of a perturbative expansion the
UV divergences simply persist. Second, in the non-planar diagrams, they tend
to ``mix'' with IR divergences \cite{MinwallaSeiberg:1999}. We refer to \cite{DouglasNekrasov:2001,Szabo:2001} for classical reviews of the topic, and to 
\cite{Steinacker:2011ix} for insights into the relation between fuzzy spaces
and matrix models.\par
Langmann, Szabo and Zarembo (LSZ) introduced and studied a scalar
field theory on the Moyal plane, and showed that its partition function
admits a matrix model representation \cite{LSZ1,LSZ2}. The LSZ matrix model
has the explicit form 
\begin{equation}
\mathcal{Z}_{\mathrm{LSZ}}\left( E,\tilde{E}\right) =\int \mathcal{D}M%
\mathcal{D}M^{\dagger }~\exp \left( -N\tr\left\{ MEM^{\dagger }+M^{\dagger }%
\tilde{E}M+\widehat{V}\left( M^{\dagger }M\right) \right\} \right) ,
\label{eq:CSMM}
\end{equation}%
where $M$ is a $N\times N$ complex matrix, $E,\tilde{E}$ are external
matrices with real eigenvalues determined by the model and $\widehat{V}$ is
a polynomial potential with quadratic and quartic interaction terms.
Recently, a Hermitian matrix model with external field $E$ and quartic
interaction has been studied in the context of NC scalar field theories \cite{deJong:2018,Grosse:2019a,Grosse:2019b}. It is worthwhile to stress that,
despite the similarities, the fact that the LSZ model is a complex random matrix ensemble leads to a different analysis and very
different set of results. In particular, we will establish a relationship
between this model and a family of matrix models that appear in Chern--Simons
theory and are related to $q$-deformed random matrix ensembles.\par
We will also consider a supersymmetric extension of the LSZ model (sLSZ), which we define as the supermatrix model
\begin{equation}  \label{eq:superLSZ}
\mathbf{Z}_{\mathrm{sLSZ}} \left( E , \tilde{E} \right) = \int \mathcal{D} M \mathcal{D}
M^{\dagger} ~ \exp \left( - (N_1 + N_2) \Str \left\{ M E M^{\dagger} + M^{\dagger} 
\tilde{E} M + \widehat{V} \left( M^{\dagger} M \right) \right\} \right) ,
\end{equation}
where now $M$ is a $(N_1 + N_2) \times (N_1 + N_2) $ complex supermatrix and $\Str$ is the supertrace. 
The details are given in section \ref{sec:super}.\par

On the other hand, a number of different matrix models have been studied in
gauge theory, more precisely in the study of certain topological and
supersymmetric gauge theories in compact three-manifolds such as Seifert manifolds.
The simplest case is that of $\cS^{3}$, where the partition function of $U(N)
$ Chern--Simons theory, admits the expression \cite{Marino:2002fk} 
\begin{equation}
\mathcal{Z}_{\mathrm{CS}}=\int_{\mathbb{R}^{N}}\prod_{1\leq j<k\leq N}\left(
2\sinh \left( \frac{x_{j}-x_{k}}{2}\right) \right) ^{2}\prod_{j=1}^{N}e^{-%
\frac{1}{2g_{s}}x_{j}^{2}}\mathrm{d}x_{j},  \label{eq:sinhCS}
\end{equation}%
with $g_{s}$ a coupling constant, which can be related to the level $k\in \Z$
of Chern--Simons theory by $g_{s}=\frac{2\pi \I }{N+k}$. A review of early
results is \cite{Marino:2005book}. The matrix model description can also be
obtained and further understood by using different types of localization of
the path integral \cite{BeasleyWitten:2005,BlauThompson:2006,Kallen:2011}.\par

Interestingly, expressions such as \eqref{eq:sinhCS}, or generalizations
thereof, with Schur polynomial insertions, describing then Wilson loop
observables in the Chern--Simons theory, are completely solvable, even for
finite $N$, using standard random matrix theory methods \cite%
{Tierz:2002,DolivetTierz:2006,RomoTierz:2011}. We will give an
interpretation of the LSZ matrix model in terms of probabilities in the
random matrix description of the observables of $U(N)$ Chern--Simons theory
on $\cS^{3}$. This type of probability is a classical object in random
matrix theory \cite{Mehta:2004} and typically computed via a Fredholm determinant. For
this family of models, such probabilities have also been studied, 
in the context of what is known as critical
statistics \cite{muttalib1993new,Nishigaki:1999} (statistics interpolating between Poisson and Wigner--Dyson behaviour). However, we will be naturally lead to the consideration of
atypically large fluctuations instead  (like for the GUE model \cite{DeanMajumdar}, only the model \eqref{eq:sinhCS} is a $q$-deformation of the GUE) rather than any bulk spectral quantity. In addition, we shall see that
different insertions of external matrices $E,\tilde{E}$ will be related to
different Chern--Simons observables.\par

Further examples beyond the topological $U(N)_{k}$ Chern--Simons theory are
the ABJ(M) theories \cite{ABJM,ABJ}, supersymmetric $%
U(N_{1})_{k}\times U(N_{2})_{-k}$ Chern--Simons-matter theories preserving twelve supercharges. The
partition function of ABJ theory on $\cS^{3}$ has the matrix model
representation \cite{Kapustin:2009kz}%
\begin{equation}
\begin{aligned} \mathbf{Z}_{\mathrm{ABJ}} = \int_{\mathbb{R}^{N_1} }
\int_{\mathbb{R}^{N_2}} & \frac{ \prod_{1 \le j<k \le N_1 } \left( 2\sinh
\left( \frac{ x_{j}-x_{k} }{2} \right) \right) ^{2} \prod_{ 1 \le r<s \le
N_2 } \left( 2\sinh \left( \frac{w_{r}-w_{s}}{2}\right) \right) ^{2} }{
\prod_{j =1} ^{N_1} \prod_{r=1} ^{N_2} \left( 2 \cosh \left( \frac{ x_j -
w_{r}}{2} \right) \right)^2 } \\ & \times \prod_{j=1}^{N_1}
e^{-\frac{1}{2g_{s}} x_{j}^{2} } \mathrm{d}x_{j} \prod_{r=1}^{N_2}
e^{-\frac{1}{2g_{s}} w _{r}^2}\mathrm{d}w_{r} , \end{aligned}
\label{eq:ABJM}
\end{equation}%
where the relation between the string coupling $g_{s}$ and the Chern--Simons
level $k\in \mathbb{Z}$ is $g_{s}=\frac{2\pi \mathrm{}i}{k}$. It was shown in 
\cite{MaPutExact} that equation \eqref{eq:ABJM} defines a supermatrix model
related to the ordinary matrix model for $U(N)$ Chern--Simons theory on the lens
space $\mathbb{S}^3 / \mathbb{Z}_2$ \cite{Aganagic:2002,Halmagyi:2003}. The approach of \cite{MaPutExact}
has been applied in \cite{ABJexact} to evaluate exactly the ABJ partition
function. We will show how the relation uncovered between LSZ and
Chern--Simons matrix models extends to the supermatrix models 
\eqref{eq:superLSZ}-\eqref{eq:ABJM}.\par
\medskip

This paper has two parts. In the first one, we will dwell on the interpretation of the
LSZ model in noncommutative field theory because, since \cite{LSZ1,LSZ2}, a closely related matrix model to \eqref{eq:CSMM} has appeared in other works,
dealing with gauge theories in noncommutative $\R^{3}$ \cite{GereJuricWallet:2015,Wallet:2016}. 
Thus, it is interesting to study further this, more so given the new
interpretations we will discuss in the second part paper of this paper. \par
The analysis of Abelian gauge theories in noncommutative $\R^{3}$ has been
carried out in \cite{VitaleWallet:2012,MartinettiVitaleWallet:2013,GereVitaleWallet:2013,Vitale:2014,GereJuricWallet:2015,Wallet:2016}. 
The deformation of three-dimensional Euclidean space considered is the
so-called $\R_{\lambda }^{3}$, modelled on the $\la2$ algebra \cite{OriginalCSandR3}. 
The discrete eigenvalues of the quadratic Casimir of $\la2$ 
produce a ``foliation'' by fuzzy spheres
of different radii. In \cite{GereVitaleWallet:2013,GereJuricWallet:2015} it
has been found the most general action satisfying three conditions: gauge
invariance, stability of the vacuum, positivity. The second condition is
tightly related to the choice of fundamental field for the theory; in
particular, two different approaches, namely expanding a covariant gauge
field around a gauge invariant flat connection \cite{GereVitaleWallet:2013}
or around the null configuration \cite{GereJuricWallet:2015} lead to
dissimilar pictures. In both cases, after gauge fixing, the gauge theories
reduce to a tower of scalar theories with quartic interaction, one on each
fuzzy sphere foliating $\R_{\lambda }^{3}$. On one hand, the first model
yields a partition function which appears to be related to a string model
with a background $B$-field enforcing string noncommutativity \cite{AlekseevRecknagelSchomerus:2000}. On the other hand, the similarity
of the action in the latter case with the Langmann--Szabo--Zarembo model 
\cite{LSZ1,LSZ2} was highlighted in \cite{GereJuricWallet:2015}. We will
show that, from \cite{GereJuricWallet:2015,Wallet:2016} but with a slight
change in the definition of the exterior calculus on $\mathbb{R}_{\lambda}^{3}$ and eventually taking the large radius limit, one reproduces the
matrix model \eqref{eq:CSMM}.\par
In the second part, which is the core of the work, we will establish the
relationship between the two sets of matrix models described above, exactly
in the manner explained. This is a random matrix result, linking two
families of models: complex matrix models with external sources on one hand, 
$q$-ensembles that appear in Chern--Simons theory on the other. Importantly, the
relationship is not between the same observables. On one hand, we have
partition functions (albeit generalized via different choices of external
matrices) and on the other hand probabilities in the $q$-ensembles, corresponding to different
observables in the Chern--Simons theory. This result can also be appreciated
independently of the gauge theoretic origin of both sets of models. The
Chern--Simons matrix model for example, also plays a prominent role in the subject of non-intersecting Dyson Brownian motion \cite{Grabiner:1999,BaikSuidan:2007}, see \cite{HaroTierz:2004,Takahashi:2012}.\par
\medskip

The article is organized as follows. A field-theoretical background for the matrix model of interest is provided in
section \ref{sec:LSZscalar}, where we review the derivation of the LSZ
matrix model from a noncommutative scalar theory. In section \ref{sec:fuzzytoLSZ} we discuss how the same matrix model emerges as a
limit in a different noncommutative field theory, the scalar theory obtained
from reduction of a three-dimensional gauge theory in $\R^3 _{\lambda}$.\par
After that, we analyze the LSZ matrix model and explain its close relation with the
Chern--Simons matrix model in section \ref{sec:CSesLSZ}. We emphasize that the connection holds in a
generalized sense, as the external matrices are not restricted to come from
a kinetic operator of a scalar field theory, and different spectra of the
external matrices will be interpreted in terms of different observables in
Chern--Simons theory. In subsection \ref{sec:fluctuations} we show that the
generalized LSZ partition function encodes certain probabilities in the
random matrix description of topological invariants computed from
Chern--Simons theory. Then, in section \ref{sec:super}, we extend the
analysis to the sLSZ supermatrix model, and establish an analogous relation
between the sLSZ generalized partition function and observables in ABJ
theory. These are the central results of the present work. Finally, the
Appendix \ref{app:Schur} contains technical details about Schur polynomials.

\section{Noncommutative scalar theory with background field}
\label{sec:LSZscalar}

In this section we review the LSZ model: in the first subsection, the geometric construction of the Moyal plane in the presence of a background magnetic field is sketched, while the second subsection is dedicated to the construction of a scalar field theory with quartic interaction. 
This provides a motivation and a physical background for the study of the matrix model \eqref{eq:CSMM}, that will be thoroughly analyzed in section \ref{sec:CSesLSZ}.\par 
The content of this section follows \cite{LSZ2}, although for the derivation of the matrix model in subsection \ref{sec:LSZMMderivation} we use a slightly different formalism than the original work, that will allow us a more direct comparison with the results in the next section. 

\subsection{Moyal plane with magnetic field}

The Moyal plane is defined through the commutation relations 
\begin{equation*}  \label{eq:MPCR}
\left[ q^{j} , q^{k} \right] = \I \theta \epsilon^{jk} ,
\end{equation*}
where $\theta$ is the essential parameter of the theory, with dimension of
length squared. A Moyal plane can always be seen as a harmonic system, in
the sense that passing to dimensionless complex coordinates one has $\left[z, \bar{z} \right] = 1$.
The noncommutative plane needs not to arise from a modification of space-time.
In fact, an example is given by a particle moving on a plane with a magnetic
field of intensity $B$ in the transverse direction; the momentum space then
becomes noncommutative $\R^2$, as the momentum operators modify according to: 
\begin{equation*}
p_j \mapsto P_j := p_j - \frac{1}{2} B \epsilon_{jk} q^{k} .
\end{equation*}
In this case the covariant momenta $P_j$ satisfy the commutation relation $\left[ P_j , P_k \right] = - \I  B \epsilon_{jk} $.

If the two frameworks are put together, that is, a transverse magnetic field
is plugged in on a noncommutative plane, three possible harmonic oscillator
pictures arise:

\begin{enumerate} [(i)]

\item on the two-dimensional position space, with annihilation and creation
operators given by the complex coordinates as above;

\item on the two-dimensional momentum space, with annihilation and creation
operator defined analogously;

\item a pair of canonical harmonic oscillators, one on each phase space
plane.
\end{enumerate}

However, the most suitable choice for us is none of them, and we will take a
mixture of all these ingredients to form two commuting copies of
annihilation and creation operators, in such a way that the problem
decouples into two one-dimensional harmonic systems. To do so, define: 
\begin{equation*}
\begin{aligned} z & := \frac{q^{1} + \I q^{2} }{\sqrt{ 2 \theta}} , \quad
\bar{z} := \frac{q^{1} - \I q^{2} }{\sqrt{2 \theta}} , \\ v & := \frac{ p_1
+ \I p_2 }{\sqrt{2 \theta^{-1}}} , \quad \bar{v} = \frac{ p_1 - \I p_2
}{\sqrt{2 \theta^{-1}}} , \end{aligned}
\end{equation*}
and use them to introduce the operators: 
\begin{equation*}
\begin{aligned} a_1 & = \frac{ z + \I v }{\sqrt{2}} , \quad a_1 ^{\dagger} =
\frac{ \bar{z} - \I \bar{v} }{\sqrt{2}} , \\ a_2 & = \frac{ \bar{z} + \I
\bar{v} }{\sqrt{2}} , \quad a_2 ^{\dagger} = \frac{ z - \I v }{\sqrt{2}} .
\end{aligned}
\end{equation*}
Straightforward calculations provide: 
\begin{equation*}
\begin{aligned} \left[a_{\alpha} , a_{\beta} ^{\dagger} \right] & =
\delta_{\alpha \beta} , \\ \left[a_{\alpha} , a_{\beta} \right] & = 0 =
\left[a_{\alpha} ^{\dagger} , a_{\beta} ^{\dagger} \right] , \end{aligned}
\end{equation*}
for $\alpha, \beta =1,2$, hence we got a pair of decoupled harmonic oscillators.\par
\begin{remark}
Lifting the obstruction $\theta$ shifts the canonical symplectic
structure on the cotangent bundle. It turns out that such shifted 2-form is
still symplectic. One can then rotate to Darboux coordinates so that the new
symplectic structure on the phase space $T^{\ast} \R^2$ is block-diagonal.
The calculations above are precisely the explicit change of coordinates.
\end{remark}\par
\medskip 
Consider now the differential operator $D_j$ associated to the
covariant momenta $P_j$, and $\tilde{D}_j$ analogous but carrying
a reflected magnetic field $-B$. If we take the arbitrary combination $-
\sigma D^2 - \tilde{\sigma} \tilde{D} ^2$ and evaluate it at the symmetric
point $\sigma = \tilde{\sigma} = \frac{1}{2}$, we obtain: 
\begin{equation*}
\left( - \sigma D^2 - \tilde{\sigma} \tilde{D} ^2 \right)_{\sigma = \tilde{%
\sigma} = \frac{1}{2} } = \vert \vec{p} \vert ^2 + \frac{B^2}{4} \vert \vec{q} \vert^2 = \theta^{-1}
\left( \frac{B^2 \theta^2}{4} \left\{ z , \bar{z} \right\} + \left\{ v , 
\bar{v} \right\} \right) ,
\end{equation*}
where the curly bracket in the right-hand side stands for anticommutation. On the
other hand, in terms of the harmonic oscillators description, we have: 
\begin{equation*}
\sum_{\alpha=1}^{2} a_{\alpha} ^{\dagger} a_{\alpha} = \frac{1}{2} \left(
\left\{ z , \bar{z} \right\} + \left\{ v , \bar{v} \right\} - \mathrm{i} [ v
, \bar{z} ] + \mathrm{i} [z, \bar{v} ] \right) ,
\end{equation*}
which means 
\begin{equation}
\left( - \sigma D^2 - \tilde{\sigma} \tilde{D} ^2 \right)_{\sigma = \tilde{%
\sigma} = \frac{1}{2} } = \frac{2}{\theta} \sum_{\alpha =1} ^2 \left(
a_{\alpha} ^{\dagger} a_{\alpha} + \frac{1}{2} \right)
\label{eq:MPkineticop}
\end{equation}
at points $B^2 \theta^2 = 4$. The preferred curves $\frac{B^2 \theta^2}{4}=1$ correspond to the self-dual
points of the Langmann--Szabo symmetry \cite{LSduality}. The theory is
independent of the actual choice of curve in parameter space we restrict to,
namely $B= \pm 2 \theta^{-1}$. In fact, the two theories we obtain are
equivalent in the Seiberg--Witten sense, i.e., they transform into the same
theory \cite{LSduality}. The invariance reflects the fact that the operators $D_j , \tilde{D}_j$ only differ by a reflection $B \mapsto -B$, thus the symmetric choice $\sigma = \tilde{\sigma}$ drops the dependence on the sign of the magnetic field.

\subsection{LSZ model}
\label{sec:LSZMMderivation}

Given a scalar field $\Phi$ on the Moyal plane, we can expand it in terms of
the Landau basis, consisting of eigenstates of both harmonic oscillators,
as: 
\begin{equation*}
\Phi = \sum_{\ell_1 , \ell_2 = 1 } ^{\infty} M_{\ell_1 \ell_2 } \lvert
\ell_1 , \ell_2 \rangle .
\end{equation*}
This expression naturally defines an infinite matrix $M$ associated to the
field $\Phi$. Now recall the kinetic operator in \eqref{eq:MPkineticop};
using the property 
\begin{equation*}
a^{\dagger} _{\alpha} a_{\alpha} \lvert \ell_1 , \ell_2 \rangle = \left(
\ell_{\alpha} -1 \right) \lvert \ell_1 , \ell_2 \rangle , \qquad
\ell_{\alpha} = 1,2, \dots, \ \alpha=1,2 ,
\end{equation*}
it is possible to write: 
\begin{equation*}
\begin{aligned} \left( \sum_{\alpha =1} ^2 \left( a_{\alpha} ^{\dagger}
a_{\alpha} + \frac{1}{2} \right) \right) \Phi & = \sum_{\ell_1 , \ell_2}
M_{\ell_1 \ell_2 } \left\{ \left( \ell_1 - \frac{1}{2} \right) + \left(
\ell_2 - \frac{1}{2} \right) \right\} \lvert \ell_1 , \ell_2 \rangle \\ & =
\sum_{\ell_1 , \ell_2 } \left( \ell_1 - \frac{1}{2} \right) \left\{
M_{\ell_1 \ell_2} \lvert \ell_1 , \ell_2 \rangle + M_{\ell_2 \ell_1 } \lvert
\ell_2 , \ell_1 \rangle \right\} . \end{aligned}
\end{equation*}
Therefore one obtains: 
\begin{equation}
\begin{aligned} \Phi^{\dagger}  \left( \sum_{\alpha =1} ^2 \left(
a_{\alpha} ^{\dagger} a_{\alpha} + \frac{1}{2} \right) \right) \Phi 
& = \sum_{\ell_1, \ell_2} \left( \ell_1 - \frac{1}{2} \right) \left\{
M^{\dagger} _{\ell_2 \ell_1} M _{\ell_1 \ell_2} + M_{\ell_2 \ell_1 }
M^{\dagger} _{\ell_1 \ell_2} \right\} \\ & = \tr \left\{ M^{\dagger} E M + M
E M^{\dagger} \right\} , \label{eq:psikappapsi} \end{aligned}
\end{equation}
where in the last line we have introduced the diagonal matrix 
\begin{equation}  \label{eq:extmatrix}
E_{\ell_1 \ell_2} = \left( \ell_1 - \frac{1}{2} \right) \frac{ 4 \pi }{N}
\delta_{\ell_1 \ell_2} .
\end{equation}\par
Consider the action \cite{LSZ1,LSZ2} 
\begin{equation*}
\begin{aligned} S_{\mathrm{LSZ}} \left( \Phi, \Phi^{\dagger} \right) = \int_{\R^2
_{\theta} } & \left\{ \frac{1}{2} \Phi^{\dagger} \left( - \sigma D^2 -
\tilde{\sigma} \tilde{D}^2 \right) \Phi + \frac{1}{2} \Phi \left( - \sigma
D^2 - \tilde{\sigma} \tilde{D}^2 \right) \Phi^{\dagger} \right. \\ & \left.
+ m_0 ^2 \Phi^{\dagger} \Phi + \frac{g_0 ^2}{2} \left( \Phi^{\dagger} \Phi
\right)^2 \right\} . \end{aligned}
\end{equation*}
Evaluated at the symmetric point $\sigma = \tilde{\sigma} = \frac{1}{2}$ it
becomes: 
\begin{equation}
\label{eq:SLSZfinal}
S_{\mathrm{LSZ}} \left( \Phi, \Phi^{\dagger} \right) = N \tr \left\{ M^{\dagger} E M
+ M E M^{\dagger} + \widehat{m}^2 M^{\dagger} M + \frac{\widehat{g}^2}{2}
\left( M^{\dagger} M \right)^2 \right\}
\end{equation}
where we have introduced the dimensionless couplings 
\begin{equation*}
\widehat{m}^2 = \left( \frac{2 \pi \theta}{N} \right) m_0 ^2 , \qquad 
\widehat{g}^2 = \left( \frac{2 \pi \theta}{N} \right) g_0 ^2 .
\end{equation*}
We have also used equations \eqref{eq:MPkineticop} and \eqref{eq:psikappapsi}, and the matrix $E$ defined in \eqref{eq:extmatrix}. In section \ref{sec:CSesLSZ} we will solve the matrix model with action \eqref{eq:SLSZfinal}.\par

Notice that, to regularize the integral, we truncate the
matrix $M$ to its top-left $N \times N$ block, which introduces a finite
cutoff at short distance $\sqrt{ \frac{2 \pi \theta}{N} }$. The full theory
is recovered in the large $N$ limit. As expected from general features of
noncommutative field theory (see e.g. \cite{Szabo:2001}), the original noncommutativity of the phase
space is eventually encoded in the noncommutativity of matrix
multiplication. Consistently, the space-time integral becomes a trace.

\section{Scalar theory on the fuzzy sphere}
\label{sec:fuzzytoLSZ}

We now discuss another noncommutative scalar field theory in two dimensions, this time on the fuzzy sphere \cite{Madore:1991}, that leads to the LSZ matrix model \eqref{eq:CSMM}.\par
Scalar theories on the fuzzy
sphere have been extensively studied, mainly looking at them as regularized
UV/IR mixing-free versions of commutative theories \cite{DolanOConnor:2001,DolanOConnor:2002,Martin:2004}. In the standard setting
the kinetic sector prevents the integration of the angular degrees of freedom, and thus to reduce to an integral over eigenvalues. A
perturbative approach in the kinetic term was proposed in \cite{OConnorSamann:2007a,OConnorSamann:2007b}, equivalent to a high-temperature
expansion, and the presence of phase transitions, together with a triple point,
was suggested. This procedure was later generalized to $\mathbb{CP}^{N}$ \cite%
{Samann:2010}. See \cite{Panero:2006} for a review,
describing both theoretical predictions and numerical results, paying
special attention to the phase transition. An extended scheme,
allowing both a high-temperature (large-interaction) and small-interaction
analysis was presented in \cite{Polychronakos:2013}, providing further
understanding of the phase structure of the model. Other nonperturbative
aspects are still under investigation: in \cite{HataTsu:2017,HataTsu:2018}
the behaviour of correlation functions of the matrix model in the disordered
phase is analyzed\footnote{For NC gauge theories on the fuzzy sphere, a procedure to reduce the partition function to a matrix model was put forward in \cite{Steinacker:2003sd}.}.\par
However, all these works lead to matrix models different from \eqref{eq:CSMM}. 
Instead, we will start with a gauge theory in a noncommutative version of $\mathbb{R}^3$ which is ``foliated'' by fuzzy spheres \cite{OriginalCSandR3}. 
The projection of the gauge theory onto each fuzzy sphere gives a noncommutative scalar theory which reduces to LSZ in the large radius limit. 
We stress that the model we consider was introduced in \cite{GereJuricWallet:2015,Wallet:2016} and it has the distinctive feature 
that it is not directly defined as a scalar theory on the fuzzy sphere, but it is derived in a roundabout way. 
The identification with the LSZ model at large radius is possible precisely because the kinetic operator of the scalar theory descends from a three-dimensional gauge theory, instead of being given by the adjoint action of the generators of $\mathfrak{su} (2)$, as the standard kinetic term for a scalar theory on the fuzzy sphere would be.\par
The rest of the section is dedicated to the proof of this equivalence. 
We first give a very brief review of the properties of the
deformation of $\R^3$ known as $\R^3 _{\lambda}$. 
Then, the second subsection is dedicated to the construction of the gauge theory
in $\R^3 _{\lambda}$, which is essentially a review of the setup of \cite{GereJuricWallet:2015,Wallet:2016} with minor modifications.
Eventually, we reduce to a scalar theory on a fuzzy sphere of fixed radius and study the
large radius limit.

\subsection{From $\R^3 _{\lambda}$ to the fuzzy sphere}

We start with a noncommutative version of the three-dimensional Euclidean
space, imposing the coordinates to satisfy the commutation relations of $\la2$. Such space is known in the literature as $\R^3
_{\lambda} $. Then, as we will see, irreducible representations of $\la2$
determine a foliation of $\R^3 _{\lambda}$ in terms of fuzzy spheres. 
We refer to \cite{OriginalCSandR3,Vitale:2014} for detailed
insights in $\R^3 _{\lambda}$.\par
Coordinates in $\R^3 _{\lambda}$ are the generators $\left\{ x^{\mu} \right\}_{\mu=1} ^{ \ 3}$ of $\la2$, up to a length scale factor $\lambda$, satisfying: 
\begin{equation*}
\left[ x^{\mu} , x^{\nu} \right] = \I \lambda \epsilon^{\mu \nu \rho} x^{\rho} .
\end{equation*}
Irreducible representation of $\la2$ in terms of $N \times N$ matrices are
labelled by non-negative half-integers $n$, with $N=(2n+1)$. Whenever $n$ is fixed the Casimir relation
implies: 
\begin{equation*}
\vec{x}^2 := \sum_{\mu=1} ^{3} \left( x^{\mu} \right)^2 = \lambda^2 n(n+1) ,
\end{equation*}
which corresponds to pick a sphere of fixed radius $r^2 = \lambda^2 n (n+1)$, denoted by $\cS^2 _n$. 
One could interpret this construction of the fuzzy
sphere as analogous to the usual embedding of $\cS^2$ into $\R^3$, but
replacing coordinates with noncommutative Hermitian operators.\par
It is well-known that the large $N$ limit of the fuzzy sphere at fixed
radius gives the commutative sphere. Conversely, the scaling limit $r \to
\infty$ keeping the parameter $\theta = \frac{r^2}{n}$ fixed leads to the Moyal plane. 
We focus on this latter setting and write: 
\begin{equation*}
\lambda^2 = \frac{ \theta}{n+1} , \quad r^2 = \theta n , \ \qquad \theta~\text{fixed}.
\end{equation*}
As shown in \cite{Vitale:2014,GereVitaleWallet:2013,GereJuricWallet:2015},
there exists a matrix basis on each fuzzy sphere, which allows to identify
fields in $\R^3 _{\lambda}$ with a tower of matrices of increasing size.
Noncommutativity is then encoded in $N \times N$ matrix multiplication, with 
$N=2n+1$, at each level $n \in \frac{1}{2} \N$.\par
Integration over $\R^3 _{\lambda}$ is defined as
\begin{equation*}
\int_{\R^3_{\lambda}} F = 2 \pi \sum_{n \in \frac{1}{2} \N} \lambda^3 (n+1) %
\tr  f^{(n)}  = 2 \pi \theta^{3/2} \sum_{n \in \frac{1}{2} %
\N} \frac{1}{\sqrt{n+1}} \tr  f^{(n)}  ,
\end{equation*}
where $f^{(n)}$ is the $N \times N$ matrix representation of the function $F$ at level $n$. 
We have had to adapt the prescription of \cite{Vitale:2014} to treat $\lambda$ and $r$ as functions of $n$. 
This integral has the expected behaviour at large $N$, giving the
volume of a sphere of radius $\sim \sqrt{\theta n}$. Freezing the radial degree of freedom, the integral reduces to 
\begin{equation*}
\int_{\cS^2 _n } F = 2 \pi \theta \tr f^{(n)} 
\end{equation*}
on the $n$-th fuzzy sphere.

\subsection{Gauge theory setup}

At this point, we introduce the algebra of derivations on $\R^3 _{\lambda}$. We define the forms 
\begin{equation*}
\tau_{\mu} := \frac{x^{\nu}}{r \lambda} \delta_{\nu \mu} , 
\end{equation*}
which yield associated derivations defined as
\begin{equation}
\label{eq:R3Lderivat}
\partial_{\mu} := \mathrm{ad}_{\tau_{\mu}} = \I \left[ \tau_{\mu} , \cdot \right] ,
\end{equation}
satisfying the ring relation 
\begin{equation*}
\left[ \partial_{\mu} , \partial_{\nu} \right] = - \frac{1}{r} \epsilon_{\mu \nu \rho} \partial_{\rho} .
\end{equation*}
Our derivations only differ from \cite{GereVitaleWallet:2013,GereJuricWallet:2015,Wallet:2016} by a factor $\lambda/r$, which belongs to the centre of the algebra and does not change the relevant properties of $\partial_{\mu}$.\par
It is also possible to introduce the 1-form $\Theta$ such that $\Theta \left( \partial_{\mu} \right) = - \I \tau_{\mu} $. It satisfies
\begin{equation*}
\mathrm{d} \Theta \left( \partial_{\mu} , \partial_{\nu} \right) + \left[ \Theta \left(
\partial_{\mu} \right), \Theta \left( \partial_{\mu} \right) \right] = 0 ,
\end{equation*}
implying that $\Theta$ is a flat connection. The most
general gauge-invariant action in this framework includes a term coupling
the gauge field to $\Theta$. However, in our analysis, it will follow from the definition \eqref{eq:R3Lderivat} that such term would only
provide a constant correction to the mass term, hence we will reabsorb it in
the definition of bare mass. This is one of the small differences between the present setting and \cite{GereJuricWallet:2015,Wallet:2016}.\par
To construct an Abelian gauge theory on $\R^3_{\lambda}$, we need to consider the noncommutative analogue of an $\mathrm{ad} (P)$-bundle, for $P \to \R^3$ a principal $U(1)$-bundle. 
This is given by the right module $\mathfrak{u}(1) \otimes \R^3 _{\lambda}$ over $\R^3 _{\lambda}$. 
We introduce the anti-Hermitian gauge fields $A_{\mu}$ and the gauge covariant derivative
\begin{equation*}
\nabla _{\mu} = \partial_{\mu} + A_{\mu} .
\end{equation*}
The curvature $F^{A}$ has components: 
\begin{equation*}
F^{A} _{\mu \nu} = \left[ \nabla_{\partial_{\mu}} , \nabla_{\partial_{\nu}} \right] - \nabla_{\left[
\partial_{\mu}, \partial_{\nu} \right]}  = \partial_{\mu} A_{\nu} - \partial_{\nu} A_{\mu} + \left[
A_{\mu} , A_{\nu} \right] + \frac{1}{r} \epsilon_{\mu \nu \rho} A_{\rho} .
\end{equation*}
We see that the curvature has the usual form plus an extra term inherited from the underlying noncommutativity. 
The group of gauge transformations is the group of unitary transformations of $\R^3 _{\lambda}$ (see \cite{GereJuricWallet:2015} and references therein), and the gauge field transforms in the usual way:
\begin{equation*}
	A_{\mu} \mapsto U ^{\dagger} A_{\mu} U + U^{\dagger} \partial_{\mu} U,
\end{equation*}
for any unitary $U$.
Furthermore, one can also consider the $\Theta$-covariant derivative $\widetilde{\nabla}$ associated to the 1-form $\Theta$, which for any field $f$ on $\R^3 _{\lambda}$ satisfies:
\begin{equation}
\label{eq:tildenablainvariant}
	\widetilde{\nabla}_{\mu} f = \partial_{\mu} f - \I \tau_{\mu} f  = - \I f \tau_{\mu} ,
\end{equation}
where the second equality follows from the definition \eqref{eq:R3Lderivat} of the derivation as a commutator with $\tau_{\mu}$. $\widetilde{\nabla}$ is gauge invariant:
\begin{equation*}
	\widetilde{\nabla}_{\mu} f \mapsto U^{\dagger} \circ \widetilde{\nabla}_{\mu} \left( U f \right) = - \I  f \tau_{\mu} ,
\end{equation*}
where we used \eqref{eq:tildenablainvariant}. Besides, using again \eqref{eq:R3Lderivat} we see that $\I \tau_{\mu}$ transforms under gauge transformations as
\begin{equation*}
	\I \tau_{\mu} \mapsto \I U^{\dagger} \tau_{\mu} U - U^{\dagger} \partial_{\mu} U .
\end{equation*}\par
For any choice of gauge fields $A_{\mu}$ we can define the so-called covariant coordinates $C_{\mu}$, defined as the components of the 1-form $\nabla - \widetilde{\nabla}$,
\begin{equation*}
C_{\mu} = \nabla_{\mu} - \widetilde{\nabla} _{\mu} = A_{\mu} + \I \tau _{\mu} .
\end{equation*}
The $C_{\mu}$ fields are clearly anti-Hermitian and behave under gauge transformations according to:
\begin{equation*}
	\left( \nabla_{\mu} - \widetilde{\nabla}_{\mu} \right) f  \mapsto U ^{\dagger} A_{\mu} U f  + \I U ^{\dagger} \tau_{\mu} U f ,
\end{equation*}
that is, $C_{\mu} \mapsto U ^{\dagger} C_{\mu} \circ U$. This is because the $U^{\dagger} \partial_{\mu} U$ parts cancel between the two contributions with opposite sign. 
On the other hand, by the very definition of derivations, see \eqref{eq:R3Lderivat}, it stems that
\begin{equation*}
     U^{\dagger} i \tau_{\mu} U = U^{\dagger} \partial_{\mu} U + \I \tau_{\mu} ,
\end{equation*}
which implies
\begin{equation}
\label{eq:CmuTransfRule}
    C_{\mu} \mapsto \left( U ^{\dagger} A_{\mu} U + U^{\dagger} \partial_{\mu} U \right) + \I \tau_{\mu} .
\end{equation}
Therefore the gauge transformations leave unchanged the $\tau_{\mu}$-part and modify the gauge field $A_{\mu}$ in the ordinary way. In particular, we underline for later convenience that, since we are working in a matrix basis in which $x^3$ is diagonal, the field $C_3$ has, by its very definition, the eigenvalues of $\tau_3$ multiplied by $\sqrt{-1}$ in its diagonal, plus the diagonal entries of $A_3$. The off-diagonal entries of $C_3$ in the basis we are working are simply the off-diagonal entries of $A_3$. 
Therefore, a gauge transformation acting on $C_3$ transforms the contribution by $A_3$ in the standard way without affecting the eigenvalues of $\tau_3$ in the diagonal of $C_3$.\par
\medskip
At this point, two options disclose to set up an Abelian gauge theory: we
ought to chose either $A_{\mu}$ or $C_{\mu}$ as the fundamental fields of the theory.
They differ by a flat connection: we expect the two resulting theories to be
related by a redefinition of the vacuum. The two different procedures have been carried out, respectively, in \cite{GereVitaleWallet:2013} and \cite{GereJuricWallet:2015}. 
We follow the second approach, taking $C_{\mu}$ as
variables, and we pursue the most general action such that:

\begin{enumerate}[(i)]
\item it is gauge invariant and at most quartic in the fundamental variable;
\item it does not involve tadpoles at classical order, which imposes not to include linear terms in the fundamental variable;
\item it is positive definite.
\end{enumerate}
The second condition is essential in order to have a stable vacuum. Taking
such family of gauge-invariant actions and writing down the classical
equations of motion, one finds out that the absolute minimum is the null configuration $C_{\mu} = 0$. 
Other local minima are present. However, due to the special form of the
derivations in the present work, all the vacua in the centre of the noncommutative algebra are constant
configurations. (Nevertheless, some other vacuum configurations may fall outside the centre of the algebra).\par
The action which satisfies the above mentioned three hypotheses is:
\begin{equation}
\label{eq:SCbeforefix}
    S \left[ C \right] = \frac{2}{g_0 ^2} \int_{ \R^3 _{\lambda} } \left( \left[ C_{\mu} , C_{\nu} \right]^2 + \Omega \left( C_{\mu}  C_{\nu} + C_{\nu} C_{\mu} \right)^2 - m_0 ^2 C^2 \right) 
\end{equation}
We rewrite it in terms of Hermitian fields as $C_{\mu} = \I \Phi_{\mu}$ and remove the redundant degree of freedom fixing the gauge  $\Phi_3 =
\tau_3$. Notice that, according to \eqref{eq:CmuTransfRule} and subsequent discussion, this corresponds to the Coulomb gauge $A_3 = 0$ in terms of the standard gauge fields $A_{\mu}$. This procedure could be done, indeed, in a BRST-invariant fashion, introducing a ghost/anti-ghost pair $(c, \bar{c})$ with grading $(-1,+1)$ together with a St\"{u}ckelberg field $b$, with zero ghost number. We add to the action a gauge fixing term
\begin{equation}
\label{eq:gaugefix}
	\tr \left\{ b \left( \Phi_3 - \tau_3 \right) + \I \bar{c} \left[ \Phi_3 , c \right] \right\} 
\end{equation}
and integrate out the field $b$ first, enforcing the gauge fixing condition $\Phi_3 = \tau_3$. Then the second term above reduces to $\bar{c} D_3 c$, hence the ghost and anti-ghost fields decouple. 
This is not an exhaustive discussion, and we refer to \cite{GereJuricWallet:2015,Wallet:2016} for more details on the definition of a BRST coboundary operator in the present theory. The proof that the gauge fixing term \eqref{eq:gaugefix} is in fact BRST-exact with respect to the just mentioned BRST operator can be found in \cite[Eq.s (3.6)-(3.9)]{GereJuricWallet:2015}. The general formalism of BV/BRST quantization of noncommutative gauge theories is discussed in \cite{Bering:2009km}.\par
At this point, we rearrange the remaining scalar fields into complex ones $\Phi, \Phi^{\dagger}$, scaled by a factor $\sqrt{2} / g_0$, and restrict to the region of parameter space for which the potential is of the form $V \left( \Phi^{\dagger} \Phi \right)$, corresponding to the choice $\Omega = \frac{1}{3}$ in \eqref{eq:SCbeforefix}.
We finally arrive at the action (see \cite[Eq. (3.17)]{Wallet:2016} and discussion around it)
\begin{equation}
\label{eq:WalletS}
S = \int_{\R^3 _{\lambda} } \left(
\Phi^{\dagger} \hat{K} \Phi + \Phi \hat{K} \Phi^{\dagger} + \frac{%
g_0 ^2}{2} \Phi^{\dagger}  \Phi  \Phi^{\dagger}  \Phi \right) ,
\end{equation}
with kinetic operator
\begin{equation*}
\hat{K} := m_0 ^2 \mathrm{Id} + \frac{8}{3 } \tau_3 \left( \tau_3 - i \partial_3 \right).
\end{equation*}
The product in \eqref{eq:WalletS} is nothing but the matrix multiplication when the fields are represented in the natural matrix basis at each level $n$. 
The sole difference between the action \eqref{eq:WalletS} and the one in \cite{Wallet:2016} is the different scaling of the kinetic operator $\hat{K}$ in the large radius limit, analyzed in the next subsection.

\subsection{Large radius limit}

At this point, we project the system onto a single fuzzy sphere. This means
we fix a half-integer $n$ and restrict to those states spanned by the $n$-th eigenstate of the quadratic Casimir operator, related to the radial coordinate. The remaining degree of
freedom is the degeneracy at fixed $n$, labelled by $k=-n,\dots ,n$. We
recall that the radius is $\sqrt{\theta n}$, with $\theta $ to be kept fixed
at large $N$, where $N=2n+1$. Fields $\Phi$ are projected onto $N\times N$
matrices $\phi$, according to the foliation of $\R_{\lambda }^{3}$. We then
calculate the matrix elements appearing in the action \eqref{eq:WalletS}: 
\begin{equation*}
\begin{aligned} \langle k \rvert \hat{K} \phi \lvert k^{\prime} \rangle
& = m_0 ^2 \langle k \rvert \phi \lvert k^{\prime} \rangle + \frac{8}{3 r^2
\lambda^2 } \left( \langle k \rvert \left( x^{3} \right)^2 \phi \lvert
k^{\prime} \rangle + \langle k \rvert x^{3} \left[ x^3, \phi \right] \lvert
k^{\prime} \rangle \right) \\ & = \langle k \rvert \phi \lvert k^{\prime}
\rangle \left\{ m_0 ^2 + \frac{8 k }{3 \theta n} \left( 2 k -
k^{\prime} \right) \right\} . \end{aligned}
\end{equation*}
We multiply this expression by a factor $2\pi \theta $ which will arise from
the integral, and obtain: 
\begin{equation*}
2\pi \theta \langle k\rvert \hat{K} \phi \lvert k^{\prime }\rangle
=(2n+1)\langle k\rvert \phi \lvert k^{\prime }\rangle \left\{ \widehat{m}^{2}+%
\frac{16\pi }{3}\frac{k\left( 2k-k^{\prime }\right) }{n(2n+1)}\right\} ,
\end{equation*}%
where we have defined the dimensionless parameter 
\begin{equation*}
\widehat{m}^{2}:=\left( \frac{2\pi \theta }{2n+1}\right) m_{0}^{2}
\end{equation*}%
by scaling the bare mass with the cutoff induced by finite $N$. Therefore,
\begin{equation*}
\begin{aligned} 2 \pi \theta \tr \left\{ \phi^{\dagger} \hat{K} \phi +
\phi \hat{K} \phi^{\dagger} \right\} = (2n+1) \sum_{k, k^{\prime} =-N}
^N & \left\{ \langle k^{\prime} \rvert \phi^{\dagger} \lvert k \rangle
\langle k \rvert \phi \lvert k^{\prime} \rangle \left( \widehat{m}^2 + \frac{16
\pi}{3} \frac{ k \left( 2k - k^{\prime} \right) }{n (2n+1) } \right) \right.
\\ & \left. + \langle k^{\prime} \rvert \phi \lvert k \rangle \langle k
\rvert \phi^{\dagger} \lvert k^{\prime} \rangle \left( \widehat{m}^2 + \frac{16
\pi}{3} \frac{ k \left( 2k - k^{\prime} \right) }{n (2n+1) } \right)
\right\} . \end{aligned}
\end{equation*}%
Switching the labels of the dummy variables $k,k^{\prime }$ in the last
summand, we obtain 
\begin{equation*}
\begin{aligned} (2n+1) \sum_{k, k^{\prime} =-n} ^n & \left\{ \widehat{m}^2
\left( \langle k^{\prime} \rvert \phi^{\dagger} \lvert k \rangle \langle k
\rvert \phi \lvert k^{\prime} \rangle + \langle k^{\prime} \rvert \phi
\lvert k \rangle \langle k \rvert \phi^{\dagger} \lvert k^{\prime} \rangle
\right) \right. \\ & \left. + \langle k^{\prime} \rvert \phi^{\dagger}
\lvert k \rangle \langle k \rvert \phi \lvert k^{\prime} \rangle \frac{32
\pi}{3} \frac{ k^2 + {k^{\prime}}^2 - k k^{\prime} }{n (2n+1)} \right\} .
\end{aligned}
\end{equation*}%
At this point, in order to compare with the LSZ model, we need to pass from
the angular momentum to the harmonic oscillator description. This is done
using the relation $k=n-\ell $ for $\ell =0,1,\dots 2n=N-1$. We thus define
the rearranged matrix $M$ as 
\begin{equation*}
M_{\ell _{1}\ell _{2}}:=\langle n-\ell _{1}\rvert \phi \lvert n-\ell
_{2}\rangle .
\end{equation*}%
In terms of the new indices we have 
\begin{equation*}
k^{2}+{k^{\prime }}^{2}-kk^{\prime }=n^{2}-n\left( \ell _{1}+\ell
_{2}\right) +\ell _{1}^{2}+{\ell _{2}}^{2}-\ell _{1}\ell _{2} .
\end{equation*}%
Hence, denoting by $S_{n}$ the restriction of the action \eqref{eq:WalletS} to a single fuzzy sphere, we get: 
\begin{equation*}
\begin{aligned} 
S_n  = (2n+1)
\sum_{\ell_1 , \ell_2 = 0 } ^{2n} & \left\{ \hat{m}^2 \left( M^{\dagger}
_{\ell_2 \ell_1 } M_{\ell_1 \ell_2} + M _{\ell_2 \ell_1 } M^{\dagger}
_{\ell_1 \ell_2} \right) \right. \\ & \left. + M^{\dagger} _{\ell_2 \ell_1 }
M_{\ell_1 \ell_2} \frac{32 \pi }{3} \frac{ n^2 - n \left( \ell_1 + \ell_2
\right) + \ell_1 ^2 + {\ell_2 }^2 - \ell_1 \ell_2 }{n (2n+1)} \right\} \\ &
+ \left( \frac{ 2 \pi \theta }{ 2n+1} \right) \frac{ g_0 ^2 }{2} \tr \left\{
\left( M^{\dagger} M \right)^2 \right\} . \end{aligned}
\end{equation*}%
Eventually, rescaling $g_{0}^{2}$ into a dimensionless parameter
\begin{equation*}
	\widehat{g}^2 :=\left( \frac{2\pi \theta }{2n+1}\right) g_{0}^{2} 
\end{equation*}
and splitting the second line into the sum of two
terms, the action on the fuzzy sphere reads: 
\begin{equation}
S_{n}=N\tr\left\{ M^{\dagger
}E^{\prime} M+M E^{\prime} M^{\dagger }+\left( \widehat{m}^{2}+\frac{8\pi }{3}\right) M^{\dagger
}M+\frac{\widehat{g}^{2}}{2}\left( M^{\dagger }M\right) ^{2}\right\} ,
\end{equation}%
where the matrix $E^{\prime}$ coming from the kinetic operator $\hat{K}$ is
\begin{equation}
E^{\prime}_{\ell _{1}\ell _{2}}=\left( -\frac{32\pi }{3N}%
\left( \ell _{1}+\frac{1}{4}\right) \right) \delta _{\ell _{1}\ell _{2}}+%
\mathcal{O}\left( \frac{1}{N^{2}}\right) .  \label{Qspec}
\end{equation}
Therefore, we find that, in the large $N$ limit, this model coincides
with the LSZ at the self-dual point, with magnetic field scaled by a factor $- \frac{8}{3}$. 
Note that the restriction to preferred points in parameter space, both for the LSZ and the model of \cite{GereJuricWallet:2015}, 
is crucial to obtain the result. Indeed, the large amount of symmetry at these preferred points produces cancellations which do not hold in general.

\section{Solution of the matrix model}
\label{sec:CSesLSZ}

As we have seen in sections \ref{sec:LSZscalar} and \ref{sec:fuzzytoLSZ}, different noncommutative
field theories reduce to a matrix model of the form \eqref{eq:CSMM} 
\begin{equation*}
\mathcal{Z}_{\mathrm{LSZ}} \left( E,\tilde{E}\right) =\int \mathcal{D}M\mathcal{D}M^{\dagger
}~\exp \left( -N\tr\left\{ M E M^{\dagger }+M^{\dagger} \tilde{E} M+\widehat{%
V}\left( M^{\dagger }M\right) \right\} \right)
\end{equation*}%
in terms of $N\times N$ complex matrices, depending on the insertion of two
external matrices. The potential $\widehat{V} \left( M^{\dagger }M\right)  $ is a quadratic polynomial with dimensionless
coefficients 
\begin{equation}
\widehat{V}\left( M^{\dagger }M\right) =\widehat{m}^{2}\left( M^{\dagger
}M\right) +\frac{\widehat{g}^{2}}{2}\left( M^{\dagger }M\right) ^{2}.
\label{eq:potential}
\end{equation}%
We also let the external fields $E,\tilde{E}$ have arbitrary eigenvalues
which, for convenience and consistently with the noncommutative field theory setting (recall Eq.\eqref{eq:extmatrix}), we write as $\frac{4 \pi}{N} \eta _{\ell } , \frac{ 4 \pi }{N}\tilde{\eta}_{\ell} $ respectively, for $\ell =1,\dots ,N$. In particular, the
original LSZ model, discussed in section \ref{sec:LSZscalar}, corresponds to external matrices
with $\eta_{\ell} = \tilde{\eta}_{\ell}$ given by
consecutive integers plus a constant shift, see Eq. \eqref{eq:extmatrix}. The unconventional presence of $\tilde{E}$ explicitly breaks $U(N)$ symmetry, but the system is still
tractable.\par
We now reduce this matrix model to an ordinary multiple integral in terms of the spectra of
the external fields, and show that the results can be written in terms of the observables
of $U(N)$ Chern--Simons theory in $\cS^{3}$, with $q= e^{- g_s}$ real.

\subsection{General solution}
\label{sec:GenSol}

As shown in \cite{LSZ2}, we can approach the solution using the singular value decomposition of a generic $N \times N$ complex matrix $M$: 
\begin{equation*}
M = U_1 ^{\dagger} \mathrm{diag} \left( {\lambda}_1 , \dots , {\lambda}_N \right) U_2 ,
\end{equation*}
with $U_{\alpha=1,2}$ unitary matrices and $\lambda _{\ell} \ge 0$.\footnote{%
See \cite{Morris:1991} for comments on this parametrization with regards to
the more usual one in terms of eigenvalues \cite{Mehta:2004}. In any case,
this transformation is much used and very useful when studying complex
matrix models, such as the LSZ.} The measure becomes 
\begin{equation*}
\mathcal{D} M ~ \mathcal{D} M^{\dagger} = \left[ \mathrm{d} U_1 \right] %
\left[ \mathrm{d} U_2 \right] \prod_{\ell =1} ^N \mathrm{d} y_{\ell} ~
\Delta_N \left[ y \right] ^2 ,
\end{equation*}
where $\left[ \mathrm{d} U_{\alpha} \right]$ is the invariant Haar measure over 
$U(N)$, and $y_{\ell} := {\lambda}_{\ell} ^2$ and 
\begin{equation}
\label{eq:VdMdef}
\Delta_N \left[ y \right] = \prod_{1 \le \ell < \ell^{\prime} \le N} \left(
y_{\ell} - y_{\ell ^{\prime}} \right)
\end{equation}
is the Vandermonde determinant.\par
As shown in \cite{LSZ2}, the use of this transformation implies that
integrations over $U_{\alpha =1,2}\in U(N)$ decouple over the two types of
external field terms. Denoting by $Y =\mathrm{diag}\left( y_{1},\dots
,y_{N}\right) $, the angular degrees of freedom $U_{\alpha }$ can be
integrated out using the Harish-Chandra--Itzykson--Zuber (HCIZ) formula \cite{HarishChandra:1956,ItzyksonZuber:1980}: 
\begin{equation}
\int_{U(N)}\left[ \mathrm{d}U_{2}\right] \exp \left\{ -N\tr\left(
EU_{2}^{\dagger } Y U_{2}\right) \right\} =\mathcal{C}_{N}\frac{\det_{1\leq
\ell ,\ell ^{\prime }\leq N}\left( e^{-4\pi \eta _{\ell }y_{\ell ^{\prime
}}}\right) }{\Delta _{N}\left[ \eta \right] \Delta _{N}\left[ y \right] }%
,  \label{eq:hcizf}
\end{equation}%
where we have used the explicit form of the eigenvalues of $E$ and
denoted 
\begin{equation*}
\mathcal{C}_{N}:=\left( 4\pi \right) ^{-\frac{N(N-1)}{2}%
}\prod_{j=1}^{N-1}j!=\left( 4\pi \right) ^{-\frac{N(N-1)}{2}}G(N+1),
\end{equation*}%
where $G(\cdot)$ is the Barnes $G$-function.\par
Analogous expression is obtained for integration over $U_{1}$ replacing $E$ by $\tilde{E}$.
For a potential $\widehat{V}\left( M^{\dagger }M\right) $ as in %
\eqref{eq:potential}, one gets: 
\begin{equation}
\begin{aligned}
\mathcal{Z}_{\mathrm{LSZ}} \left( E,\tilde{E}\right) &= \int_{[0, \infty)^{N}} \Delta_{N}\left[ y \right] ^{2}~  \prod_{\ell=1}^{N} e^{-\left( \widehat{m}^{2} y_{\ell}+\frac{\widehat{g}^{2}}{2} y_{\ell}^{2}\right) } \mathrm{d} y_{\ell}  \\
& \times \int \left[ \mathrm{d}U_{2}\right] \exp
\left( -NEU_{2}^{\dagger } Y U_{2}\right) \ \int \left[ \mathrm{d}U_{1}\right]
\exp \left( -N\tilde{E}U_{1}^{\dagger } Y U_{1}\right) 
\label{eq:formula1}
\end{aligned}
\end{equation}%
Plugging HCIZ \eqref{eq:hcizf} into %
\eqref{eq:formula1}, we note that the partition function for the matrix
model is: 
\begin{equation*}
\mathcal{Z}_{\mathrm{LSZ}} \left( E,\tilde{E}\right) =\mathcal{C}_{N} ^2 \int_{[0, \infty)^{N}} \Delta _{N}\left[ y \right] ^{2}~ \frac{\det_{1\leq \ell ,\ell ^{\prime }\leq N}\left( e^{-4\pi \eta _{\ell } y_{\ell ^{\prime }}}\right) \ \det_{1\leq \ell ,\ell ^{\prime
}\leq N}\left( e^{-4\pi \tilde{\eta}_{\ell } y_{\ell ^{\prime }}}\right) }{%
\Delta _{N}\left[ y \right] ^{2}\Delta _{N}\left[ \eta \right] \Delta _{N}%
\left[ \tilde{\eta}\right] } ~
\prod_{\ell =1}^{N} e^{- \widehat{m}^{2} y_{\ell } - \frac{\widehat{g}^{2}}{2} y_{\ell}^{2} } \mathrm{d} y_{\ell } ,
\end{equation*}%
where we recall that $\eta _{\ell } $ (respectively $\tilde{\eta}_{\ell }$) stand for the
eigenvalues of $E$ (respectively $\tilde{E}$), up to a factor $\frac{4\pi }{N}$. 
This was already done in \cite{LSZ2} but now, after applying the HCIZ
formula, we will not expand the resulting determinants into sums of
permutations of $N$ objects, and cancel a Vandermonde squared instead. That is, after a
suitable rescaling of the integration variables: 
\begin{equation}
\mathcal{Z}_{\mathrm{LSZ}} \left( E,\tilde{E}\right) =\frac{\mathcal{C}_{N}^{\prime }}{\Delta _{N}%
\left[ \eta \right] \Delta _{N}\left[ \tilde{\eta}\right] }\int_{[0, \infty)^{N}} \det_{1\leq \ell ,\ell ^{\prime }\leq
N}\left( e^{-\eta _{\ell } y_{\ell ^{\prime }}}\right) \ \det_{1\leq \ell
,\ell ^{\prime }\leq N}\left( e^{-\tilde{\eta}_{\ell } y_{\ell ^{\prime
}}}\right) ~  \prod_{\ell
=1}^{N}e^{- m^{2} y_{\ell }-\frac{%
g^{2}}{2} y_{\ell }^{2} }~\mathrm{d} y_{\ell },  \label{eq:formula2}
\end{equation}%
with coefficients redefined as:
\begin{equation}
\label{eq:scalemg}
	m^{2}= \frac{\widehat{m}^{2}}{4\pi },\qquad g^{2}=\frac{\widehat{g}^{2}}{(4\pi )^{2}}, 
\end{equation}
and
\begin{equation*}
\mathcal{C}_{N}^{\prime }=\left( 4\pi \right) ^{-N}\mathcal{C}_{N} ^2 = \frac{ G(N+1)^2}{ (4 \pi)^{N^2} } =\frac{%
2^{-N\left( N-1\right) } \pi^{N}}{\mathrm{vol}\left( U(N)\right) ^2 },
\end{equation*}%
where $\mathrm{vol} ( \cdot )$ is the volume of the gauge group. Note that this
normalization is essentially the square of the partition function of the Gaussian unitary ensemble (GUE) \cite{Mehta:2004}.\par
In the theory of non-intersecting Brownian motion, the determinants in %
\eqref{eq:formula2} are very familiar. This whole theory of determinantal
processes is known to be directly related to $U(N)$ Chern--Simons theory on $%
\cS^{3}$ and with the Wess--Zumino--Witten (WZW) model \cite{HaroTierz:2004}%
, where such connection was shown to follow from specializations of the
determinants 
\begin{equation*}
\det_{1\leq \ell ,\ell ^{\prime }\leq N}\left( e^{-\eta _{\ell } y_{\ell
^{\prime }}}\right) ,\quad \det_{1\leq \ell ,\ell ^{\prime }\leq N}\left(
e^{-\tilde{\eta}_{\ell } y_{\ell ^{\prime }}}\right) 
\end{equation*}%
in \eqref{eq:formula2}. However, it is more direct to show the relation
through the corresponding matrix model formulation. Recall for this the
definition of a Schur polynomial \cite{Macdonaldbook} 
\begin{equation*}
s_{\mu }\left( x_{1},\dots ,x_{N}\right) =\frac{\det_{1\leq \ell ,\ell
^{\prime }\leq N}\left( x_{\ell }^{\mu _{\ell ^{\prime }}+N-\ell ^{\prime
}}\right) }{\det_{1\leq \ell ,\ell ^{\prime }\leq N}\left( x_{\ell }^{N-\ell
^{\prime }}\right) },
\end{equation*}%
and hence we rewrite the scaled eigenvalues $\eta _{\ell },\tilde{\eta}_{\ell }$ of the external matrices $E,\tilde{E}$, assuming they are
integers, as: 
\begin{equation}
 \label{eq:spectra} 
\begin{aligned}
\eta _{\ell } &=\mu _{\ell }+N-\ell , \\
\tilde{\eta}_{\ell } &=\nu _{\ell }+N-\ell , 
\end{aligned}
\end{equation}%
for $\ell =1,\dots ,N$. Without loss of generality, we can relabel the eigenvalues so that $\mu_1 \ge \mu_2 \ge \cdots \ge \mu_N \ge 0$, and likewise for $\nu_1 \ge \cdots \ge \nu_N $. Then, using 
\begin{equation*}
\prod_{1\leq \ell <\ell ^{\prime }\leq N}\left( e^{y_{\ell }}-e^{y_{\ell
^{\prime }}}\right) ^{2}=\prod_{1\leq \ell <\ell ^{\prime }\leq N}\left(
2\sinh \left( \frac{y_{\ell }-y_{\ell ^{\prime }}}{2}\right) \right)
^{2}\prod_{m=1}^{N}e^{(N-1) y_{m}}
\end{equation*}
and reflecting the variables $y_{\ell} \mapsto - y_{\ell}$, we immediately have\footnote{With our definition, $\eta_{\ell}$ and $\tilde{\eta}_{\ell}$ are non-negative integers, which is the physical choice motivated by the LSZ model. If we want them to be non-positive integers, we simply do not reflect $y_{\ell} $ and define $\eta_{\ell}, \tilde{\eta}_{\ell}$ with opposite sign. The discussion would be exactly the same, except for the integration domain being $[0, \infty)$.}: 
\begin{equation}
\begin{aligned}
\mathcal{Z}_{\mathrm{LSZ}} \left( \mu, \nu \right) = \frac{ (-1)^{N} \mathcal{C}_{N}^{\prime }}{ \Delta_N [\eta ]   \Delta_N [\tilde{\eta} ]} 
\int_{ (-\infty, 0]^{N}} & \prod_{1\leq \ell <\ell ^{\prime }\leq N}\left( 2\sinh \left( 
\frac{y_{\ell }-y_{\ell ^{\prime }}}{2}\right) \right) ^{2} \prod_{\ell =1}^{N}\mathrm{d} y_{\ell
}~e^{ \beta y_{\ell }- \frac{g^{2}}{2} y_{\ell
}^{2}  } \\ \times & s_{\mu }\left(
e^{y_{1}},\dots ,e^{y_{N}}\right) s_{\nu }\left( e^{y_{1}},\dots
,e^{y_{N}}\right) ,
\end{aligned}
\label{eq:ZLSZgenY}
\end{equation}%
with $\beta =m^{2}-N+1$. We henceforth write $\mathcal{Z}_{\mathrm{LSZ}} \left( \mu, \nu \right)$ for $\mathcal{Z}_{\mathrm{LSZ}} \left( E,\tilde{E}\right)$, stressing the dependence on the partitions $\mu =(\mu_1, \dots, \mu_N) $ and $\nu = ( \nu_1, \dots, \nu_N )$. 
Except for the integration domain, the expression \eqref{eq:ZLSZgenY} is close to the general
version of the $U(N)$ Chern--Simons on $\cS^{3}$ matrix model, with two
different insertions of Schur polynomials, whose evaluation gives the topological invariant of a pair of unknots \cite{Marino:2005book,Kimura:2015dfa} carrying the $U(N)$ representations $\mu$ and $\nu$. 
It is worthwhile to mention that the same matrix model \eqref{eq:ZLSZgenY} is also related to the Hopf link invariant, but now with representations $\mu$ and $\nu^{\ast}$, where $\nu ^{\ast }:=\left( \nu _{1}-\nu _{N},\nu _{1}-\nu _{N-1},\dots,0\right)$. We give the details in Appendix \ref{app:hopf}. 
Furthermore, if the external matrix $E$ has positive integer eigenvalues while $\tilde{E}$ has negative integers eigenvalues, or vice versa, the relation then is with the Hopf link invariant carrying representations $\mu$ and $\nu$.\par
Instead, after a shift of variables in \eqref{eq:ZLSZgenY} and using the identity \eqref{eq:schurcid}, we obtain the matrix model representation%
\begin{equation*}
\begin{aligned}
\mz_{\mathrm{LSZ}} \left( \mu ,\nu \right) =\mathcal{A}_N \left(\mu , \nu \right) \int_{ (-\infty , - \gamma ]^{N}} & \prod_{j=1}^{N}\mathrm{d}x_{j}~e^{-\frac{g^{2}}{2}%
\sum_{j=1}^{N}x_{j}^{2}}\prod_{1\leq j<k\leq N}\left( 2\sinh \left( \frac{%
x_{j}-x_{k}}{2}\right) \right) ^{2} \\ \times & s_{\mu }\left( e^{x_{1}},\dots
,e^{x_{N}}\right) s_{\nu }\left( e^{x_{1}},\dots,e^{x_{N}}\right) ,
\end{aligned}
\end{equation*}%
where $\gamma = \beta/g^2$ and
\begin{equation}
\label{eq:formula3}
\mathcal{A}_N \left(\mu, \nu \right) := \frac{ (-1)^{N} \mathcal{C}_{N}^{\prime } }{  \Delta_N [\eta ]   \Delta_N [\tilde{\eta} ] } \ \exp \left( \frac{\gamma^{2} N}{2} + \gamma \left( \vert \mu \vert + \vert \nu \vert \right) \right) .
\end{equation}\par
Notice also that the Vandermonde factors in the denominator of \eqref{eq:formula3}, which
depend exclusively on the eigenvalues of the external matrices, can be
written, using Weyl's denominator formula (see Appendix \ref{app:Schur}), as%
\begin{eqnarray*}
\Delta _{N}\left[ \eta \right] &=&\prod_{1\leq \ell <\ell ^{\prime }\leq
N}\left( \mu _{\ell }-\mu _{\ell ^{\prime }}-\ell +\ell ^{\prime }\right)
=G(N+1)\mathrm{dim}\mu , \\
\Delta _{N}\left[ \tilde{\eta}\right] &=&\prod_{1\leq \ell <\ell ^{\prime
}\leq N}\left( \nu _{\ell }-\nu _{\ell ^{\prime }}-\ell +\ell ^{\prime
}\right) =G(N+1)\mathrm{dim} \nu ,
\end{eqnarray*}%
These Barnes $G$-functions cancel in \eqref{eq:formula3} against the ones coming from the double application of HCIZ formula. We finally obtain%
\begin{equation}
\label{solLSZmunu}
\boxed{
\mathcal{Z}_{\mathrm{LSZ}} \left( \mu, \nu \right) = C_N \left(\gamma, \vert \mu \vert , \vert \nu \vert \right)  \ \frac{ \mathcal{Z}_{\mathrm{CS}} \ \left\langle W_{\mu}  W_{\nu} ~\Id_{(- \infty, - \gamma ]^{N}} \right\rangle_{\mathrm{CS}} }{ \dim \mu ~ \dim \nu } ,
}\end{equation}%
where the relation between the partitions $\mu, \nu$ and the external matrices of the generalized LSZ model is given in equation \eqref{eq:spectra}, and 
\begin{equation*}
	 C_N \left(\gamma, \vert \mu \vert , \vert \nu \vert \right) = \frac{ (-1)^{N}  e^{ \frac{\gamma^{2} N}{2} + \gamma \left( \vert \mu \vert + \vert \nu \vert \right) } }{ (4 \pi)^{N^2}  } .
\end{equation*}
In formula \eqref{solLSZmunu}, $\mathcal{Z}_{\mathrm{CS}}$ is the $U(N)$ Chern--Simons partition function on $\cS^3$,
which is a quantum topological invariant also known as
Witten--Reshetikhin--Turaev invariant \cite{Witten:1989,ReshetikhinTuraev:1990,ReshetikhinTuraev:1991}, defined as the matrix model \eqref{eq:sinhCS} and whose explicit evaluation we give below. 
Besides, $W_{\mu}$ is the trace of the holonomy of the gauge connection along an unknot inside $\mathbb{S}^3$, taken in the $U(N)$ representation corresponding to the partition $\mu$, and likewise for $W_{\nu}$. 
In \eqref{solLSZmunu}, the Chern--Simons coupling is $g_s=\frac{1}{g^{2}}$. In Chern--Simons theory, $g_{s}$ is related to the Chern--Simons level $k$ by $g_s=\frac{2\pi \mathrm{i}}{N+k}$, while
the real string coupling constant $g_{s}$ is used when describing topological strings. That is the same type of description here, since ${g}^{2}$ is real. 
Finally, $\langle \cdots \rangle_{\mathrm{CS}}$ in \eqref{solLSZmunu} means the average in the Chern--Simons matrix model \eqref{eq:sinhCS}, and $\Id_{(- \infty, - \gamma ]^{N}}$ is the $N$-dimensional indicator function
\begin{equation*}
	\Id_{(- \infty, - \gamma ]^{N}} (x_1, \dots, x_N )= \prod_{j=1} ^{N} \Id_{(- \infty, - \gamma ]} (x_j) = \begin{cases} 1, &  x_j \le - \gamma \quad \forall \ j = 1, \dots, N , \\ 0 , & \text{otherwise} .\end{cases}
\end{equation*}
Therefore, $\left\langle  W_{\mu}  W_{\nu} \Id_{(- \infty, - \gamma]^{N}} \right\rangle_{\mathrm{CS}}$ would correspond to the two-unknot invariant, but averaged only using variables $x_j \le - \gamma$ (instead of $x_j \in \R$). 
Its relation with the actual invariant of a pair of unknots is further discussed in subsection \ref{sec:fluctuations} through the lenses of random matrix theory.\par

\subsection{Quantum dimensions}

An important particular case of the above general setting is when one of the
partitions in \eqref{eq:spectra} is void. That is, in terms of LSZ theory, one has an external
matrix with the equispaced spectra and the other one generalized with a
partition, see the definitions \eqref{eq:extmatrix} and \eqref{eq:spectra}. 
This case corresponds, as we shall see, to quantum dimensions in
the Chern--Simons interpretation \cite{DolivetTierz:2006}. Quantum dimension
of a representation associated to the partition $\mu $ is given by the
following hook-content formula \cite{Macdonaldbook,DolivetTierz:2006}%
\begin{equation*}
\mathrm{dim}_{q}\mu :=\prod_{x\in \mu }\frac{\lfloor N+c(x)\rfloor _{q}}{%
\lfloor h(x)\rfloor _{q}},
\end{equation*}%
where for each box $x\equiv (j,k)$ of the Young diagram determined by $\mu $%
, the quantity $h(x):=\mu _{j}+\mu _{k}^{\prime }-j-k+1$ is the hook length,
with the prime meaning conjugate diagram, and $c(x):=j-k$ is known as the
content of the box $x$. The operation $\lfloor \cdot \rfloor _{q}$ denotes
the symmetric $q$-number, that is%
\begin{equation*}
\lfloor n\rfloor _{q}=\frac{q^{n/2}-q^{-n/2}}{q^{1/2}-q^{-1/2}}.
\end{equation*}%
In Chern--Simons theory on $\cS^3$, the unknot invariant is given by quantum dimensions \cite{Marino:2005book,DolivetTierz:2006}.
Since one of the two external matrices has harmonic oscillator spectrum the
matrix model above reduces to%
\begin{equation*}
\begin{aligned}
\mz_{\mathrm{LSZ}} \left( \mu , \emptyset \right)= \mathcal{A}_N \left( \mu , \emptyset \right) \int_{\left( -\infty ,- \gamma \right] ^{N}} & \prod_{1\leq j<k\leq N}\left( 2\sinh
\left( \frac{x_{j}-x_{k}}{2}\right) \right) ^{2} \prod_{j=1}^{N}e^{-\frac{g^{2}}{2}x_{j}^{2}} \mathrm{d}x_{j} \\ 
\times & s_{\mu }\left( e^{x_{1}},\dots ,e^{x_{N}} \right) ,  
\end{aligned}
\end{equation*}%
whose evaluation leads to%
\begin{equation}
\label{mmqdim}
\boxed{
	\mz_{\mathrm{LSZ}} \left( \mu, \emptyset \right) = C_N ( \gamma, \vert \mu \vert , 0 ) \  \frac{ \mathcal{Z}_{\mathrm{CS}} \ \left\langle W_{\mu} ~\Id_{(- \infty, - \gamma]^{N}} \right\rangle_{\mathrm{CS}} }{ \mathrm{dim} \mu } .
}\end{equation}
This should be compared with the unknot invariant computed in Chern--Simons theory, which differs from the present setting in the fact that the integral is taken over $\mathbb{R}^N$ instead of $(-\infty, - \gamma]^N$. We will come back to this point in subsection \ref{sec:fluctuations}. 
The exact evaluation of the unknot invariant gives \cite{DolivetTierz:2006}
\begin{equation*}
	\langle W_{\mu} \rangle = q^{-\frac{1}{2}C_{2}\left( \mu \right) }\mathrm{dim}_{q}\mu ,
\end{equation*}
where $q=e^{-1/g^{2}}$. 
Besides, in the expression above the term%
\begin{equation*}
C_{2}\left( \mu \right) =\left( N+1\right) \vert \mu \vert +\sum_{\ell=1} ^{N} \left( \mu_{\ell}^{2}-2\ell \mu _{\ell}\right)
\end{equation*}%
is the $U(N)$ quadratic Casimir of the representation $\mu $, labelled by the Young
diagram associated to the partition $\mu $, with $\mu _{\ell}$ boxes in the $\ell$-th row, with rows understood to be aligned on the left.\par
The appearance of quantum dimensions is interesting in that they appear as
well in the study of noncommutative gauge theories through the analysis of
WZW D-branes \cite{AlekseevRecknagelSchomerus:2000,Steinacker:2002}.
However, as we have seen in section \ref{sec:LSZscalar}, only the simpler setting, where the two external matrices are equal and
have harmonic oscillator spectra, is directly linked to a noncommutative scalar theory.

\subsection{Chern--Simons and LSZ partition function}

We have shown in section \ref{sec:LSZscalar} that our study of noncommutative scalar
field theory naturally leads to a LSZ matrix model with $E=\tilde{E}$ and
spectra $\frac{4 \pi}{N} \left( \ell - \frac{1}{2} \right)$ for $\ell =
1,...,N$, see Eq. \eqref{eq:extmatrix} (or the same but scaled and shifted, see \eqref{Qspec}). 
Thus, we consider now the case in which
both partitions are void: this corresponds to the two external matrices
having harmonic oscillator spectra. In particular, from \eqref{eq:spectra},
we have that $\eta _{\ell }=N-\ell$ for $\ell =1,...,N$. The fact that the
two physical spectra have an overall energy shift only has an impact at the level of
renormalization of the mass parameter. This follows immediately from a
simple property of Schur polynomials, given in Appendix \ref{app:Schur}.\par
Then one obtains the matrix model without Schur polynomial insertions, and
the corresponding matrix integral is related to the one for the Chern--Simons partition function, given in \eqref{eq:sinhCS} (recall that $g_{s}=1/g^2$). The Chern--Simons matrix model has the exact solution \cite{Tierz:2002}: 
\begin{equation}
\label{eq:ZCSexactSW}
\mathcal{Z}_{\mathrm{CS}}=\left( \frac{2\pi }{g^{2}}\right) ^{N/2} N! ~ e^{\frac{N(N+1)(N-1)}{6 g^2}}\prod_{j=1}^{N-1}\left( 1 - q^{j}\right)^{N-j},
\end{equation}%
with $q= e^{-1/g^2}$ as above. The product can also be written as a $q$-deformed Barnes function, which, in the limit $%
g\rightarrow \infty $ (which is $q\rightarrow 1$), reduces to the Barnes
$G$-function. Then, the LSZ matrix model partition function is%
\begin{equation}
\label{eq:LSZpf}
\boxed{
\mz_{\mathrm{LSZ}} \left( \emptyset, \emptyset \right) = \frac{ (-1)^{N} e^{ \frac{\gamma^{2} N}{2} } }{\left( 4\pi \right) ^{N^2} }  \ \mathcal{Z}_{\mathrm{CS}} \ \langle \Id_{(- \infty, - \gamma]^{N}} \rangle_{\mathrm{CS}} . 
}\end{equation}%
Notice that, apart from the simple prefactor, the ratio between the LSZ and the Chern--Simons partition function is the average $\langle \Id_{(- \infty, - \gamma]^{N}} \rangle_{\mathrm{CS}} $ of the $N$-dimensional indicator function. This aspect is further analyzed the next subsection.\par

\subsection{Probabilistic interpretation}
\label{sec:fluctuations}

The main result \eqref{solLSZmunu} and its specializations \eqref{mmqdim} and \eqref{eq:LSZpf} admit an interpretation in terms of probabilities of large deviations of the smallest or largest eigenvalue of Hermitian random matrices. 
Consider a generic weight function $\omega (x)$, and let $\mz_{\omega}$ be the associated Hermitian random matrix ensemble,
\begin{equation}
\label{eq:omegaensemble}
	\mz_{\omega} = \int_{\R^N} \Delta_{N} [x]^{2} \prod_{j=1} ^{N} \omega (x_j) ~\mathrm{d} x_j .
\end{equation}
The probability that the largest eigenvalue of a random matrix in the ensemble \eqref{eq:omegaensemble} is smaller than a given threshold $s \in \mathbb{R}$ is
\begin{equation}
\label{eq:probc}
\begin{aligned}
	\mathrm{Prob}_{\omega} (x_{\mathrm{max}} \le s ) & = \mathrm{Prob}_{\omega} ( x_1 \le s , \dots, x_N \le s )  \\
		& = \frac{1}{\mz_{\omega}} \int_{(- \infty, s ]^N} \Delta_{N} [x]^{2} \ \prod_{j=1} ^{N} \omega (x_j) \mathrm{d} x_j = \left\langle \Id_{(- \infty, s ]^{N}} \right\rangle_{\omega} ,
\end{aligned}
\end{equation}
where in the last expression $\langle \cdot \rangle_{\omega}$ means the average in the ensemble \eqref{eq:omegaensemble} and $\Id_{(- \infty, s ]^{N}}$ is the $N$-dimensional indicator function. 
In the Coulomb gas picture, replacing the weight $\omega (x)$ by $\omega (x) \Id_{(- \infty, s]} (x)$ introduces a hard wall placed at $x= s$ which leaves the charges on its left. See \cite{Chekhov} for the large $N$ limit of matrix models in presence of hard walls. For the GUE ensemble, the probabilities of large fluctuations at large $N$ have been found in \cite{DeanMajShort,DeanMajumdar}.\par
We immediately see from equation \eqref{eq:LSZpf} that
\begin{equation*}
	\frac{\mz_{\mathrm{LSZ}} ( \emptyset, \emptyset) }{ \mz_{\mathrm{CS}} } = \frac{ (-1)^{N} e^{ \frac{\gamma^2 N}{2 } } }{ \left( 4\pi \right) ^{N^2} } \ \mathrm{Prob}_{\mathrm{CS}} (x_{\mathrm{max}} \le - \gamma ) ,
\end{equation*}
hence the ratio between the partition function of the LSZ model and that of Chern--Simons theory is effectively computing the probability of large deviations of eigenvalues in the Chern--Simons ensemble \eqref{eq:sinhCS} (or, strictly speaking, in the Stieltjes--Wigert ensemble \cite{Tierz:2002}, see below), up to a completely determined, parameter-dependent overall factor. More in general, formula \eqref{solLSZmunu} states that
\begin{equation*}
	\frac{\mz_{\mathrm{LSZ}} (\mu , \nu ) }{ \mz_{\mathrm{CS}} } = C_N ( \gamma, \vert \mu \vert , \vert \nu \vert ) \  \frac{ \left\langle W_{\mu } W_{\nu} \right\rangle_{\mathrm{CS}} }{ \dim \mu  \dim \nu } \ \mathrm{Prob}_{\mathrm{CS};  W_{\mu} W_{\nu} } (x_{\mathrm{max}} \le - \gamma  ) ,
\end{equation*}
where by $\mathrm{Prob}_{\mathrm{CS};  W_{\mu} W_{\nu}} (\cdot )$ we mean the probability in the matrix ensemble computing the invariant of two unknots, that is, in the Chern--Simons ensemble \eqref{eq:sinhCS} with the insertion of two Schur polynomials. This probability must be normalized by $\mz_{\mathrm{CS}} \cdot \left\langle W_{\mu } W_{\nu} \right\rangle_{\mathrm{CS}}$ and not only $\mz_{\mathrm{CS}}$. 
Therefore, the LSZ matrix model with general assignment of the external matrices, divided by the Chern--Simons partition function, is proportional to the two-unknot invariant weighted by the probability of large deviations in the random matrix description of such topological invariant. The proportionality constant yields an elementary dependence on the size $N$ and on the free parameters of the generalized LSZ theory.\par
We emphasize that the LSZ partition function does not compute the (typically small) fluctuations of the largest eigenvalue around the edge of the eigenvalue distribution. Equivalently, in the Coulomb gas picture, the LSZ partition function does not describe the fluctuations of the rightmost charge around the equilibrium. What it gives is the probability of an atypically large fluctuation, with the greatest eigenvalue moving deep into the bulk of the eigenvalue distribution. This is a $q$-analogue of the large fluctuations in the GUE discussed in \cite{DeanMajShort,DeanMajumdar} (see also \cite{MajumdarSchehr,DharMajumdarSchehr}). Besides, we underline that, differently from \cite{DeanMajumdar} where a large deviation from the equilibrium configuration is of order $\sim \sqrt{N}$, in the $q$-analogue the support of the eigenvalue density grows as $N g_s$, thus large deviations from the equilibrium are of order $\sim N$.\par
\begin{figure}[tb]
		\centering
			\includegraphics[width=0.65\textwidth]{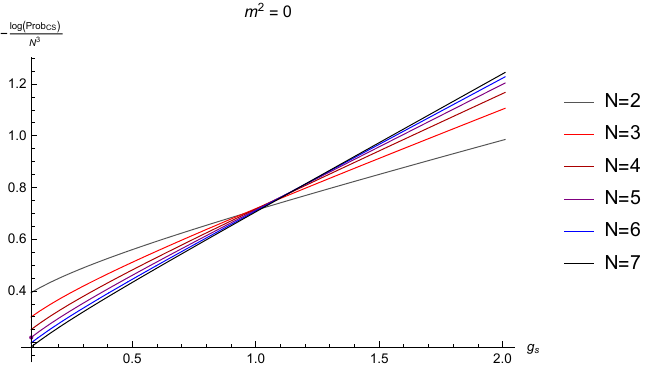}
			\caption{$-\frac{1}{N^3} \ln \frac{ \mz_{\mathrm{LSZ}}  ( \emptyset, \emptyset ) }{ \mathcal{B}_N \mz_{\mathrm{SW}} }$ as a function of the coupling $g_s$ at $m^2=0$. The curves correspond to $N$ from $2$ to $7$.}
			\label{fig:LSZvsLSW}
\end{figure}
\begin{figure}[tb]
		\centering
			\includegraphics[width=0.45\textwidth]{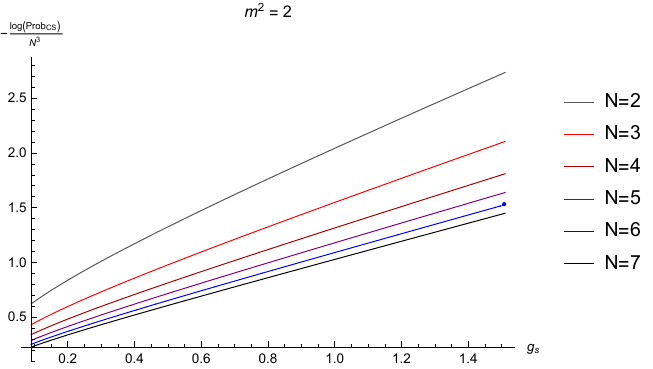}%
			\includegraphics[width=0.45\textwidth]{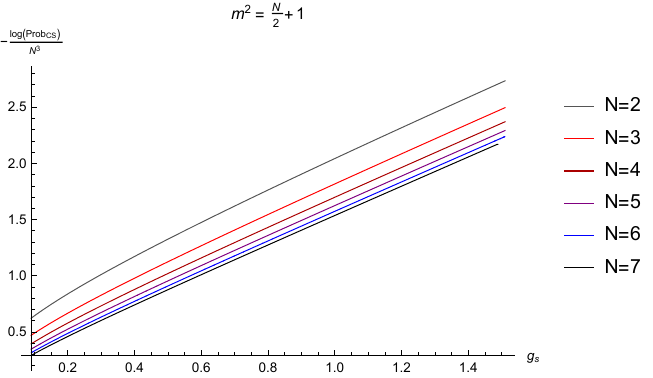}%
			\caption{$-\frac{1}{N^3} \ln \frac{ \mz_{\mathrm{LSZ}}  ( \emptyset, \emptyset ) }{ \mathcal{B}_N \mz_{\mathrm{SW}} }$ as a function of the coupling $g_s$ at $m^2=2$ (left) and $m^2 = \frac{N}{2} +1 $ (right). The curves correspond to $N$ from $2$ to $7$.}
			\label{fig:probMassive}
\end{figure}

\medskip
Now, if we consider equation \eqref{eq:formula2} without Schur insertions, $\mu=\emptyset=\nu$, then, changing variables $u_{j} = e^{y_{j} - (m^2+2N - 1)/g^2}$ \cite{Tierz:2002}, 
it is well-known that
we have a standard-random matrix ensemble with a log-normal weight, named
Stieltjes--Wigert (SW) ensemble. That is: 
\begin{equation}
\label{eq:LSZSWensemble}
	\mz_{\mathrm{LSZ}} (\emptyset, \emptyset ) = \mathcal{B}_N \int_{[s, \infty)^{N} } \prod_{1 \le j < k \le N} \left( u_j - u_k \right)^2 \prod_{j=1} ^{N} e^{ - \frac{1}{2 g_s} \left( \ln u_j \right)^2} \mathrm{d} u_j  ,
\end{equation}
for $g_s = 1/g^2$ and $s= e^{\frac{m^2 +2N -1}{g^2}}$, and with
\begin{equation*}
	\mathcal{B}_N :=  (4 \pi)^{-N^2}  \exp \left[ \frac{ N  }{2 g^2} \left( m^2 -1 \right) (m^2 +2N -1) \right] .
\end{equation*}
Note that a SW ensemble is not centered around $0$, as the weight is supported on $u \ge 0$. Therefore, up to a proportionality factor, the LSZ partition function $\mz_{\mathrm{LSZ}} (\emptyset, \emptyset)$ normalized by the SW partition function gives the probability of atypically large fluctuations of the eigenvalues away from the left edge at $u=0$, with the smallest eigenvalue deep into the bulk, $u_{\mathrm{min}} \ge s$, in a Stieltjes--Wigert ensemble:
\begin{equation*}
	\frac{ \mz_{\mathrm{LSZ}}  ( \emptyset, \emptyset ) }{ \mz_{\mathrm{SW}} } = \mathcal{B}_N  ~ \mathrm{Prob}_{\mathrm{SW}} \left( u_{\mathrm{min}} \ge s \right) .
\end{equation*}
To evaluate this quantity numerically at finite $N$, it is convenient to rewrite the integrals in the numerator and denominator of $\mathrm{Prob}_{\mathrm{SW}} \left( u_{\mathrm{min}} \ge s \right)$ as determinants \cite{Mehta:2004}, obtaining:
\begin{align*}
	\mathrm{Prob}_{\mathrm{SW}} \left( u_{\mathrm{min}} \ge s \right) & = \frac{ \det_{1 \le j,k \le N}  \left[  \int_{s} ^{\infty} u^{j+k-2} e^{ - \frac{1}{2 g_s} \left( \ln u \right)^2} \mathrm{d} u \right]    }{ \det_{1 \le j,k \le N}  \left[  \int_{0} ^{\infty} u^{j+k-2} e^{ - \frac{1}{2 g_s} \left( \ln u \right)^2} \mathrm{d} u \right]    } \\
	& = 2^{-N} \frac{ \det_{1 \le j,k \le N}  \left[  e^{\frac{g_s}{2} (j+k-1)^2 }  \mathrm{erfc} \left( \sqrt{\frac{g_s}{2}} ( m^2 +2N  -j -k ) \right) \right] }{  \det_{1 \le j,k \le N}  \left[  e^{\frac{g_s}{2} (j+k-1)^2 } \right]  } .
\end{align*}
In the latter expression, $\mathrm{erfc} (z)=1 - \mathrm{erf} (z)$ is the complementary error function. The denominator is known, and is readily extracted from Eq. \eqref{eq:ZCSexactSW}. We obtain: 
\begin{align*}
	\mathrm{Prob}_{\mathrm{SW}} \left( u_{\mathrm{min}} \ge s \right)  &= \frac{ q^{ \frac{N}{3} (2 N^2 -1) } }{ 2^{N} (1-q)^{\frac{N}{2} (N-1)} \prod_{j=1} ^{N-1} (1-q^{j} )^{N-j}  } \\
	& \times  \det_{1 \le j,k \le N}  \left[  e^{\frac{g_s}{2} (j+k-1)^2 }  \mathrm{erfc} \left( \sqrt{\frac{g_s}{2}} ( m^2 +2N  -j -k ) \right) \right] ,
\end{align*}
and we recall that $q=e^{-g_s}$ and the argument of $\mathrm{erfc}$ is related to $s$ through $\ln s = g_s (m^2 + 2N -1)$. We plot the logarithm of this probability at finite $N$ in figure \ref{fig:LSZvsLSW}. It is also clear from the definition of $s$ that the role of any fixed $m^2$ becomes less relevant as $N$ is increased, unless $m^2$ itself is increased linearly with $N$. This aspect is shown in figure \ref{fig:probMassive}.
\par
\par

We stress once more that the present setting is very different from the study of (typical) small fluctuation of the largest or smallest eigenvalue around the edge, which are suppressed by inverse powers of $N$. See \cite{DeanMajumdar,MajumdarSchehr} for thorough discussion on this point, and \cite{Vivo:2015} for further insights in the theory of large deviations. 
In the more general case, $\mz_{\mathrm{LSZ}} (\mu, \nu)$ is proportional to the probability of a large fluctuation of the eigenvalues away from the left edge in a SW ensemble deformed by the insertion of two Schur polynomials $s_{\mu} (u_1, \dots, u_N)$ and $ s_{\nu}  (u_1, \dots, u_N)$.\par

\subsection{Large $N$ limit}
In the LSZ model, the parameter $N$ regularizes the path integral of the NC field theory, as discussed in section \ref{sec:LSZscalar}. Therefore, it is natural to consider the large $N$ limit of $\mz_{\mathrm{LSZ}}  ( \emptyset, \emptyset )$, and more generally of $\mz_{\mathrm{LSZ}}  ( \mu, \nu )$. In turn, the probabilistic interpretation of the Chern--Simons theory observables also calls for a large $N$ analysis, from the perspective of large deviations theory \cite{Vivo:2015}.\par
The knowledge of the large $N$ solution of the matrix model $\mz_{\mathrm{LSZ}} (\mu, \nu )$ on either the complex matrix model side or on the side of the probability in Chern--Simons observables, would directly provide the solution on the other side. However, there are difficulties in solving the large $N$ limit from both perspectives. As mentioned in subsection \ref{sec:GenSol}, and already observed in the original work \cite{LSZ2}, a difficulty in the study of the LSZ partition function, generalized to arbitrary external matrices $E,\tilde{E}$, comes from the presence of $\tilde{E}$, which prevents a reformulation of the model in terms of a single Hermitian matrix $M^{\dagger} M$. Therefore, it becomes hard to solve the large $N$ limit of the matrix model explicitly, using for example the loop equations (that are zero-dimensional Schwinger--Dyson equations) \cite{Makeenko:1991} when $\tilde{E} \ne 0$. For $\tilde{E}=0$, the large $N$ solution has been found in \cite{LSZ2}, but this choice does not yield meaningful observables in Chern--Simons theory.\par
Taking the converse route, we could try to analyze the large $N$ asymptotics of \eqref{eq:LSZSWensemble}, adapting the argument of \cite{DeanMajumdar} to the present case. This would amount to solve a constrained extremization problem defined as follows. Write 
\begin{equation*}
	t \equiv g_s N = N/g^2 , \qquad z \equiv \log s = g_s (m^2 +2N -1)
\end{equation*}
and take the large $N$ 't Hooft limit of \eqref{eq:LSZSWensemble} keeping $t$ and $z$ fixed. Let $\rho_{\mathrm{SW}} (u)$ be the constrained SW eigenvalue density at large $N$, supported on the interval $[z, L (z)]$. We have stressed the dependence of the upper bound $L(z)$ on $z$, and let the more standard dependence on the 't Hooft coupling $t$ implicit. Then $\rho_{\mathrm{SW}} (u)$ solves the saddle point equation 
\begin{equation}
\label{eq:SPEconstrained}
	\mathrm{P} \int_z ^{L(z)} \mathrm{d} u^{\prime} ~ \frac{\rho_{\mathrm{SW}} (u^{\prime}) }{  u - u^{\prime} } = \frac{1}{2 t} \frac{ \ln u }{ u } ,
\end{equation}
with the symbol $\mathrm{P} \int$ meaning the Cauchy principal value integral. The upper boundary of the support $L(z)$ is fixed as a function of $z$ and $t$ by the normalization condition 
\begin{equation*}
	\int_z ^{L(z)} \mathrm{d} u ~ \rho_{\mathrm{SW}} (u) = 1 .
\end{equation*}
This problem cannot be solved by the method of \cite{DeanMajumdar}, because of the non-polynomial form of the right-hand side of \eqref{eq:SPEconstrained}. More precisely, the extremization problem described by the saddle point equation \eqref{eq:SPEconstrained} does not satisfy the hypothesis of Tricomi's theorem. A complete solution at large $N$ would entail an extension of the method applied in \cite{DeanMajumdar,DeanMajShort} to $q$-ensembles.\par
There are, nevertheless, two simplifying limits in which the matrix model becomes tractable at large $N$:
\begin{itemize}
	\item The $g^2 \to \infty$ limit. The interaction term in the LSZ action dominates, and the fields become non-dynamical. This corresponds to $g_s \to 0$, that is, $q \to 1$ from below. In this limit, we recover the Gaussian ensemble from the SW ensemble (up to an overall factor, readable from \eqref{eq:ZCSexactSW}), and the results of \cite{DeanMajumdar} directly apply to the present setting.
	\item The $g^2 \to 0$ limit. The quartic interaction should become tractable in standard perturbation theory in the LSZ field theory. From the Chern--Simons perspective, this corresponds to $g_s \to \infty$, equivalently $q \to 0$ from above, and simplifications take place.
\end{itemize}
In the first case, the eigenvalue density is \cite{DeanMajumdar}
\begin{equation*}
	\lim_{g_s \to 0^{+} } \rho_{\mathrm{SW}} (u) = \frac{1}{2 \pi} \sqrt{ \frac{ L(z) - u }{ u-z } } \left[ L(z) -z + 2 u \right] ,
\end{equation*}
with $L(z) = \frac{2}{3} \sqrt{z^2 +6} + \frac{1}{3} z $, and now $z \equiv \gamma / \sqrt{N}$. From \cite[Eq. (58)]{DeanMajumdar} and a change of variables $x \mapsto \sqrt{2 g_s} x$ we have 
\begin{align*}
	& \ln \mathrm{Prob}_{\mathrm{CS}} (x_{\mathrm{max}} \le -\gamma ) = \ln \mathrm{Prob}_{\mathrm{CS}} (x_{\mathrm{min}} \ge \gamma ) \\
	& \approx \frac{N^2}{2} \ln \frac{g_s}{2} - \frac{N^2}{ 54} \left[  36 z^2 - z^4 +z (z^2 +15) \sqrt{ z^2 +6 } + 27 \left[ \ln 18 -2 \ln  \left( \sqrt{z^2 +6} -z \right) \right] \right] 
\end{align*}
in the $N \to \infty$ and $g_s \to 0$ limit. This formula together with \eqref{eq:LSZpf} describes the large $N$ limit of the LSZ model in the strong interaction regime $g^2 \to \infty$.\par
The converse limit $g_s \to \infty$, that is, $g^2 \to 0$, can be analyzed as well. Changing variables $x \mapsto \sqrt{2 g_s} x$ in the CS ensemble and using 
\begin{equation*}
	2 \sinh \left(  \sqrt{\frac{ g_s}{2}} (x_j - x_k ) \right)  \approx  \exp \left( \sqrt{\frac{ g_s}{2}} \lvert x_j - x_k \rvert \right)
\end{equation*}
when $g_s \to \infty$, we get 
\begin{equation*}
	\mz_{\mathrm{LSZ}} ( \emptyset, \emptyset) =  \frac{ (-1)^{N} e^{ \frac{\gamma^2 N}{2 } } }{ \left( 4\pi \right) ^{N^2} } \ \left( \frac{g_s}{2} \right)^{\frac{N^2}{2} } \int_{(- \infty, - \gamma]^{N}} \prod_{j=1} ^{N}\mathrm{d}x_j ~ \exp \left[ - \sum_{j=1}^{N} x_j ^2 + \sqrt{\frac{g_s}{2}} \sum_{j \ne k} \lvert x_j - x_k \rvert  \right] .
\end{equation*}
The latter expression describes the partition function of a constrained one-dimensional Coulomb gas \cite{Cunden1,Cunden2}, corresponding to placing a hard wall at $x=- \gamma$ in a jellium model \cite{Baxter:63}. Here we derive the large $N$ asymptotics at leading order for this constrained model.\par
Assuming $x$ grows as $x= \xi N^{\alpha}$ in the large $N$ limit, for some $\alpha>0$ and fixed $\xi$, the first term in the exponential grows as $N^{1 + 2 \alpha}$, while the second term grows as $N^{2 + \alpha}$. A non-trivial saddle point exists for $\alpha=1$. Therefore, the large $N$ limit in this large $g_s$ regime is governed by the eigenvalue density $\rho_0 (\xi )$ that solves the saddle point equation 
\begin{equation*}
	 \sqrt{\frac{g_s}{2}} \int_z ^{L (z)} \mathrm{d} \xi^{\prime} ~ \rho_0 (\xi ^{\prime} ) \mathrm{sign} \left( \xi - \xi ^{\prime} \right) =  \xi ,
\end{equation*}
with, this time, $z= \gamma/ N$, for the scaling of $\gamma$ with $N$ to be consistent with the growth of the eigenvalues. 
Splitting the integral in two pieces, with $\xi ^{\prime} <\xi$ and $\xi ^{\prime}>\xi $ respectively, and taking the derivative of the saddle point equation in the interior of the domain, $z<\xi <L(z)$, we find
\begin{equation*}
	\rho_0 (\xi) = \frac{1}{\sqrt{2 g_s}} .
\end{equation*}
The normalization condition then imposes 
\begin{equation*}
	\int_z ^{L(z)}  \mathrm{d} \xi~ \rho_0 (\xi ) =1 \ \Longrightarrow \ L(z) -z = \sqrt{2 g_s} .
\end{equation*}
With this eigenvalue density we obtain 
\begin{equation*}
	\ln (-1)^{N} \mz_{\mathrm{LSZ}} ( \emptyset, \emptyset) \approx - N^3 \left[ z^2 + z \sqrt{2 g_s}  + \frac{g_s}{3} \right]
\end{equation*}
at leading order in $N$ and in the large $g_s$ regime.\par
The complete solution will then interpolate between this two limiting situations. The $N^2$-behaviour will correspond to small 't Hooft coupling, $t \to 0$, while letting $t $ grow linearly with $N$ will give back the $N^3$-behaviour.

\subsection{Energy levels of the generalized LSZ model}

To conclude this section, we briefly comment on the interpretation of the generalized result \eqref{solLSZmunu} in terms of the LSZ noncommutative scalar field theory \cite{LSZ1,LSZ2}. 
The absence of a preferred temporal direction in the Moyal plane prevents us from a meaningful Hilbert space picture of 
the partition function \eqref{eq:LSZpf}, and of its generalization \eqref{solLSZmunu} as well. 
However, we can still provide an
interpretation in terms of energy density levels. The system in facts splits
into two clearly separated (on the LSZ side) contributions: a Landau Hamiltonian density and
a quartic interaction term. Throughout the solution, the former is encoded in the
Vandermonde determinants, together with the Schur polynomials in the most general case, while the interaction appears as the Gaussian
measure in \eqref{eq:sinhCS}.\par
In the pure LSZ case, with both external matrices with harmonic oscillator
spectra, the Landau levels are $\frac{2 \ell - 1}{\theta}$, being $\theta$ the noncommutativity parameter (recall the details from section \ref{sec:LSZscalar}).
Introducing equal partitions $\mu = \nu$ to generalize the external matrices
would correspond to a distortion of the Landau spectrum. As an example, the
insertion of an antisymmetric partition $\mu = \left( 1, \dots, 1 \right)$
shifts the whole spectra by one level in the negative direction; as shown in
Appendix \ref{app:Schur}, this corresponds to a renormalization of the mass. 
The symmetric partition $\mu= \left( N, 0, \dots, 0 \right)$ determines the
same spectrum, although obtained by taking the highest energy level and
sending it to the bottom. Moreover, the triangular partition $\mu = \left(
N-1, N-2, \dots, 1, 0 \right) $ sends all the Landau levels to the lowest
one. More in general, any triangular partition with rows with
decreasing number of boxes from $N-n_0$ to $1$, $1 \le n_0 \le N$, and the
remaining rows void, introduces a cutoff at the $n_0$-th Landau
level and projects all the
higher energy states onto it. We underline that such cutoff, determined by the choice of partition $\mu$, is intrinsically different from the naturally induced short-distance cutoff $\sqrt{\frac{2 \pi \theta}{N}}$.

\section{Solution of the supermatrix model}
\label{sec:super}

The goal of the present section is to extend the analysis of the LSZ matrix model developed in section \ref{sec:CSesLSZ} to the supermatrix model
\begin{equation*}
\mathbf{Z}_{\mathrm{sLSZ}} \left( E , \tilde{E} \right) = \int \mathcal{D} M \mathcal{D} M^{\dagger} ~ \exp \left( - (N_1+N_2) \Str \left\{ M E M^{\dagger} + M^{\dagger}  \tilde{E} M + \widehat{V} \left( M^{\dagger} M \right) \right\} \right) 
\end{equation*}
introduced in Eq. \eqref{eq:superLSZ}. The integral is over complex supermatrices of size $(N_1 + N_2) \times (N_1 + N_2) $, and the external supermatrices $E, \tilde{E}$ have $N_1 + N_2 $ real eigenvalues each, which we write as $\frac{4 \pi}{N_1 + N_2 } \left( \eta_{\ell}, \xi_{r} \right)$ and $\frac{4 \pi}{N_1 + N_2} \left( \tilde{\eta}_{\ell}, \tilde{\xi}_{r} \right)$ for $E$ and $\tilde{E}$ respectively. We use indices $\ell=1, \dots, N_1$ and $r= 1, \dots, N_2$. 
The potential $\widehat{V}$ has a quadratic and a quartic interaction, as in \eqref{eq:potential}. We write the partition functions of ordinary matrix models with curly $\mathcal{Z}$ and supermatrix models with bold $\mathbf{Z}$.\par
In the next subsection we present a few generalities about supermatrices, then we will extend the derivation of section \ref{sec:CSesLSZ} to the sLSZ supermatrix model, this time establishing a connection with ABJ(M) theory.

\subsection{Supermatrix models}
We now introduce supermatrices and supermatrix models \cite{Yost:1991,AlvarezGaume:1991}. We consider the general case of $(N_1 + N_2) \times (N_1 +N_2)$ supermatrices, which can be defined in the block form
\begin{equation*}
	M = \left( \begin{matrix} A & \psi \\ \chi & B \end{matrix} \right) ,
\end{equation*}
where $A$ and $B$ are respectively $N_1 \times N_1$ and $N_2 \times N_2$ complex matrices with bosonic entries, and $\psi$ and $\chi$ are respectively $N_2 \times N_1$ and $N_1 \times N_2$ matrices with fermionic entries. The supermatrix $M$ is acted on by the unitary supergroup $U( N_1 \vert N_2)_{\text{left}} \times U( N_1 \vert N_2)_{\text{right}}$. 
The supertrace operation is
\begin{equation*}
	\Str M = \tr A - \tr B ,
\end{equation*}
and the integration measure $\mathcal{D} M \mathcal{D} M^{\dagger}$ is the product of Haar measures
\begin{equation*}
	\mathcal{D} M \mathcal{D} M^{\dagger} = \mathcal{D} A \mathcal{D} A^{\dagger}  \mathcal{D} B \mathcal{D} B^{\dagger}  \mathcal{D} \psi \mathcal{D} \chi .
\end{equation*}\par
We will need the supersymmetric version of the Harish-Chandra--Itzykson--Zuber formula \eqref{eq:hcizf}, which reads \cite{Alfaro:1994,Guhr:1996}
\begin{equation}
\label{eq:susyHCIZ}
\begin{aligned}
	& \int_{U(N_1 \vert N_2) }  [\mathrm{d} U]  \exp \left\{ -(N_1+N_2) \Str \left( E U Y U \right) \right\} \\
	& = \mathbf{C}_{N_1 N_2} \frac{ \prod_{\ell=1} ^{N_1} \prod_{r=1} ^{N_2} ( y_{\ell} - z_r  ) }{ \Delta_{N_1} [y] \Delta_{N_2} [z] } \frac{ \prod_{\ell=1} ^{N_1} \prod_{r=1} ^{N_2} ( \eta_{\ell} - \xi_r  ) }{ \Delta_{N_1} [\eta] \Delta_{N_2} [\xi] } 
   \det_{1 \le \ell, \ell^{\prime} \le N_1} \left( e^{- 4 \pi  \eta_{\ell } y_{\ell^{\prime}} } \right) \det_{1 \le r , r^{\prime} \le N_2 } \left( e^{4 \pi  \xi_{r} z_{r^{\prime}} } \right) ,
\end{aligned}
\end{equation}
where $E$ and $Y$ are supermatrices with eigenvalues $\frac{ 4 \pi}{N} ( \eta_{\ell}, \xi_{r} ) $ and $(y_{\ell}, z_r)$ respectively, for $\ell = 1, \dots, N_1$ and $r=1, \dots, N_2$, and the coefficient is
\begin{equation*}
	\mathbf{C}_{N_1 N_2} = \frac{ (4 \pi)^{N_1 N_2} }{ (4 \pi)^{ \frac{N_1 (N_1 -1)}{2} }  (- 4 \pi)^{ \frac{N_2 (N_2 -1)}{2} } } G (N_1 + 1 ) G (N_2 + 1 ) ,
\end{equation*}
where $G(\cdot)$ is as usual the Barnes $G$-function. The factors $\Delta_N$ are Vandermonde determinants, introduced in \eqref{eq:VdMdef}. We note the appearance of the terms
\begin{equation*}
	\frac{ \Delta_{N_1} [y] \Delta_{N_2} [z] }{ \prod_{\ell=1} ^{N_1} \prod_{r=1} ^{N_2} ( y_{\ell} - z_r  ) } ,
\end{equation*}
and likewise for $(\eta, \xi)$, which is the superdeterminant (called Berezinian) version of the Vandermonde. When $N_2=0$, formula \eqref{eq:susyHCIZ} reduces to the well known HCIZ formula \eqref{eq:hcizf}.

\subsection{General solution}
We now focus on the supermatrix model $\mathbf{Z}_{\mathrm{sLSZ}}$ defined in \eqref{eq:superLSZ}, and solve it as we have done in section \ref{sec:CSesLSZ} for the LSZ model.\par
We first rewrite the partition function as\footnote{The Jacobian is the squared Vandermonde Berezinian, see \cite[Sec. 4]{Yost:1991} and references therein.}
\begin{equation*}
\begin{aligned}
\mathbf{Z}_{\mathrm{sLSZ}} \left( E,\tilde{E}\right) &= \int_{[0, \infty)^{N_1 }} \int_{[0, \infty)^{N_2 }}  \frac{ \Delta_{N_1}\left[ y \right] ^{2}  \Delta_{N_2}\left[ z \right] ^{2} }{ \prod_{\ell=1}^{N_1} \prod_{r=1} ^{N_2} (y_{\ell} - z_r)^2 }~ \prod_{\ell=1}^{N_1} e^{- \widehat{m}^{2} y_{\ell}  - \frac{\widehat{g}^{2}}{2}  y_{\ell}^{2}  } \mathrm{d} y_{\ell}  \prod_{r=1}^{N_2}  e^{\widehat{m}^{2} z_r  + \frac{\widehat{g}^{2}}{2}  z_r ^{2}  }  \mathrm{d} z_{r}  \\
& \times \int_{U(N_1 \vert N_2) } \left[ \mathrm{d}U_{2}\right] \exp
\left( -NEU_{2}^{\dagger } Y U_{2}\right) \ \int_{U(N_1 \vert N_2) } \left[ \mathrm{d}U_{1}\right]
\exp \left( -N\tilde{E}U_{1}^{\dagger } Y U_{1}\right) ,
\end{aligned}
\end{equation*}%
and then apply the supersymmetric HCIZ formula \eqref{eq:susyHCIZ} twice. As in section \ref{sec:CSesLSZ}, we simplify the Jacobian with the denominator coming from \eqref{eq:susyHCIZ}, and obtain
\begin{equation*}
\begin{aligned}
	\mathbf{Z}_{\mathrm{sLSZ}} \left( E,\tilde{E}\right) &= \mathbf{C}_{N_1 N_2} ^2  \frac{ \prod_{\ell=1}^{N_1} \prod_{r=1} ^{N_2} ( \eta_{\ell} - \xi_{r} ) ( \tilde{\eta}_{\ell} - \tilde{\xi}_{r} )  }{ \Delta_{N_1} [ \eta ] \Delta_{N_1} [ \xi ] \Delta_{N_2} [ \tilde{\eta} ] \Delta_{N_2} [ \tilde{\xi} ] } \\
	& \times \int_{[0, \infty)^{N_1 }} \prod_{\ell=1} ^{N_1} e^{ - \widehat{m}^2 y_{\ell}   - \frac{\widehat{g}^2}{2}  y_{\ell} ^2  } \mathrm{d} y_{\ell}  ~  \det_{1 \le \ell, \ell^{\prime} \le N_1 } \left( e^{- 4 \pi  \eta_{\ell } y_{\ell^{\prime}} } \right) \det_{1 \le \ell, \ell^{\prime} \le N_1 } \left( e^{- 4 \pi  \tilde{\eta}_{\ell } y_{\ell^{\prime}} } \right) \\
	& \times \int_{[0, \infty)^{N_2 }} \prod_{r=1} ^{N_2}  e^{ \widehat{m}^2  z_r +\frac{\widehat{g}^2}{2}  z_r ^2  } \mathrm{d} z_{r} ~ \det_{1 \le r, r^{\prime} \le N_2} \left( e^{4 \pi  \xi_{r} z_{r^{\prime}} } \right) \det_{1 \le r, r^{\prime} \le N_2} \left( e^{4 \pi \tilde{\xi}_{r} z_{r^{\prime}}  } \right) .
\end{aligned}
\end{equation*}
We now assume the eigenvalues of the external supermatrices $E, \tilde{E}$ are $\frac{4 \pi}{N_1 + N_2}$ times integers, and we rewrite them in the form
\begin{equation}
\label{eq:superspectra}
\begin{aligned}
\eta _{\ell } & =\mu_{1; \ell }+N_1-\ell , \qquad \xi_{r} = \mu_{2; r}+N_2-r  \\
\tilde{\eta}_{\ell } & =\nu_{1;\ell }+N_1-\ell , \qquad \tilde{\xi}_{r} = \nu_{2;r}+N_2-r ,
\end{aligned}
\end{equation}
which is the obvious extension of \eqref{eq:spectra}. In our conventions, the partitions with subindex $1$ have rows labelled by $\ell=1, \dots, N_1$ and partitions with subindex $2$ have rows labelled by $r=1, \dots, N_2$. We again recognize the Schur polynomials,
\begin{equation*}
\begin{aligned}
	\mathbf{Z}_{\mathrm{sLSZ}} \left( E,\tilde{E}\right) &= \mathbf{C}_{N_1 N_2} ^{\prime}   \frac{ \prod_{\ell=1}^{N_1} \prod_{r=1} ^{N_2} ( \eta_{\ell} - \xi_{r} ) ( \tilde{\eta}_{\ell} - \tilde{\xi}_{r} )  }{ \Delta_{N_1} [ \eta ] \Delta_{N_1} [ \xi ] \Delta_{N_2} [ \tilde{\eta} ] \Delta_{N_2} [ \tilde{\xi} ] } \\
	& \times \int_{[0, \infty)^{N_1 }} \Delta_{N_1} [e^{y}]^2 ~ s_{\mu_1} ( e^{-y_1}, \dots, e^{-y_{N_1}} ) s_{\nu_1} ( e^{-y_1}, \dots, e^{-y_{N_1}} ) \prod_{\ell=1} ^{N_1}  e^{ - m^2 y_{\ell}  - \frac{g^2}{2} y_{\ell} ^2 } \mathrm{d} y_{\ell}  \\
	& \times \int_{[0, \infty)^{N_2 }} \Delta_{N_2} [e^{z}]^2 ~ s_{\mu_2} ( e^{z_{1}}, \dots, e^{z_{N_2}} ) s_{\nu_2} ( e^{z_{1}}, \dots, e^{z_{N_2}} ) \prod_{r=1} ^{N_2} e^{ m^2 z_r  + \frac{g^2}{2} z_r ^2 } \mathrm{d} z_{r} 
\end{aligned}
\end{equation*}
where we defined the parameters $m$ and $g$ as in \eqref{eq:scalemg} to reabsorb the factor $4 \pi$, and
\begin{equation*}
	 \mathbf{C}_{N_1 N_2} ^{\prime} = (4 \pi)^{-(N_1 + N_2) }  \mathbf{C}_{N_1 N_2} ^{2} = \frac{  G (N_1 + 1 )^2 G (N_2 + 1 )^2 }{ (4 \pi)^{(N_1 -N_2)^2 } } .
\end{equation*}
We see that the present setting closely resembles what we obtained in section \ref{sec:CSesLSZ}, now with two sets of integration variables and two pairs of Schur polynomials. 
In fact, the integrals over the two sets of variables are factorized, and we may give the result as a product of two copies of the result in section \ref{sec:CSesLSZ}.\par
However, we follow a different path and assemble the two pairs of Schur polynomials into two supersymmetric Schur polynomials labelled by representations of the supergroup $U(N_1 \vert N_2)$, see \cite{Berele:1987yi,SUSYSchur} for definitions and properties. 
Irreducible $U(N_1 \vert N_2)$ representations are classified in typical and atypical, and here we need the typical ones. The two typical representations that appear in our computations correspond to the Young diagrams
\begin{equation*}
\begin{aligned}
	\mu = (\kappa +\mu_1 ) \sqcup \lambda  , \\
	\nu = (\tilde{\kappa} + \nu_1 ) \sqcup \tilde{\lambda} ,
\end{aligned}
\end{equation*}
where $\kappa$ and $\tilde{\kappa}$ are rectangular $N_1 \times N_2$ diagrams. This means that the Young diagram $\mu$ is composed by a rectangle $\kappa$, a Young diagram $\mu_1$ on the right of it and another diagram $\lambda$ below it, and analogously for the Young diagram $\nu$. 
For these typical representations, the supersymmetric Schur polynomials decompose as \cite{Berele:1987yi,SUSYSchur}
\begin{equation*}
\begin{aligned}
	S_{\mu} ( e^{y_1}, \dots, e^{y_{N_1}} \vert e^{z_1}, \dots, e^{z_{N_2}} )  = s_{\mu_1} (e^{y_1}, \dots, e^{y_{N_1}}) s_{\lambda^{\prime}} (e^{z_1}, \dots, e^{z_{N_2}} ) \prod_{\ell=1}^{N_1} \prod_{r=1} ^{N_2} \left( e^{y_{\ell}} + e^{z_r}  \right) , \\
	S_{\nu} ( e^{y_1}, \dots, e^{y_{N_1}} \vert e^{z_1}, \dots, e^{z_{N_2}} )  = s_{\nu_1} (e^{y_1}, \dots, e^{y_{N_1}}) s_{\tilde{\lambda}^{\prime}} (e^{z_1}, \dots, e^{z_{N_2}} ) \prod_{\ell=1}^{N_1} \prod_{r=1} ^{N_2} \left( e^{y_{\ell}} + e^{z_r}  \right) ,
\end{aligned}
\end{equation*}
where $S_{\mu}$, $S_{\nu}$ are the supersymmetric Schur polynomials and $\lambda^{\prime}, \tilde{\lambda}^{\prime}$ are the conjugate partitions to $\lambda, \tilde{\lambda}$. 
Therefore, identifying the generic $\lambda, \tilde{\lambda}$ to be $\mu_2 ^{\prime}$ and $\nu_2 ^{\prime}$ in our case, and following the same manipulations as in section \ref{sec:CSesLSZ}, we rewrite $\mathbf{Z}_{\mathrm{sLSZ}}$ as
\begin{equation}
\label{eq:finalZSLSZ}
\begin{aligned}
	\mathbf{Z}_{\mathrm{sLSZ}} \left( \mu, \nu \right) &= \mathbf{A}_{ N_1, N_2} \left( \mu, \nu \right) \times \int_{(-\infty, - \gamma_1]^{N_1}} \int_{(\gamma_1, \infty]^{N_2}}  \prod_{j=1} ^{N_1}  e^{ - \frac{g^2}{2} x_{j} ^2  } \mathrm{d} x_{j}  \prod_{r=1} ^{N_2} e^{ \frac{g^2}{2} w_{r} ^2  }  \mathrm{d} w_{r} \\
	& \times \frac{ \prod_{1 \le j < k \le N_1 }  \left( 2 \sinh \left( \frac{ x_{j} - x_{k}}{2} \right) \right)^2 ~ \prod_{1 \le r<s \le N_2 }  \left( 2 \sinh \left( \frac{ w_{r} - w_{s}}{2} \right) \right)^2   }{ \prod_{j=1} ^{N_1} \prod_{r=1} ^{N_2}  \left(2 \cosh \left( \frac{ x_j - w_{r} }{2} \right) \right)^2 }  \\
	& \times S_{\mu} ( e^{x_1}, \dots, e^{x_{N_1}} \vert e^{w_1}, \dots, e^{w_{N_2}} ) S_{\nu} ( e^{x_1}, \dots, e^{x_{N_1}} \vert e^{w_1}, \dots, e^{w_{N_2}} ) ,
\end{aligned}
\end{equation}
with $\gamma_{\alpha} = m^2 -N_{\alpha} +1$, $\alpha=1,2$, and the overall coefficient being
\begin{equation*}
\begin{aligned}
	 \mathbf{A}_{ N_1, N_2} \left( \mu, \nu \right)  & = (-1)^{N_1} \mathbf{C}_{N_1 N_2} ^{\prime} \frac{ \prod_{\ell=1}^{N_1} \prod_{r=1} ^{N_2} ( \eta_{\ell} - \xi_{r} ) ( \tilde{\eta}_{\ell} - \tilde{\xi}_{r} )  }{ \Delta_{N_1} [ \eta ] \Delta_{N_1} [ \xi ] \Delta_{N_2} [ \tilde{\eta} ] \Delta_{N_2} [ \tilde{\xi} ] } \\
	 & \times  \exp \left( \frac{\gamma_1 ^2 N_1 - \gamma_2 ^2 N_2}{2 } + \gamma_1 \left( \vert \mu_1 \vert + \vert \nu_1 \vert \right) + \gamma_2 \left( \vert \mu_2 \vert + \vert \nu_2 \vert \right) \right) \\
	&=  (-1)^{N_1} \frac{ \prod_{\ell=1}^{N_1} \prod_{r=1} ^{N_2} ( \eta_{\ell} - \xi_{r} )^2 ( \tilde{\eta}_{\ell} - \tilde{\xi}_{r} )^2 }{ (4\pi)^{(N_1-N_2)^2} \dim\mu ~ \dim\nu } \\
	& \times \exp \left( \frac{\gamma_1 ^2 N_1 - \gamma_2 ^2 N_2}{2 } + \gamma_1 \left( \vert \mu_1 \vert + \vert \nu_1 \vert \right) + \gamma_2 \left( \vert \mu_2 \vert + \vert \nu_2 \vert \right) \right) .
\end{aligned}
\end{equation*}
For the second equality, we have written the contribution from the eigenvalues of the external supermatrices $E, \tilde{E}$ in terms of the partitions $\mu, \nu$, and noted again that the Barnes $G$-functions coming from the dimensions of the representations cancel with those arisen from the supersymmetric HCIZ formula, precisely as in the ordinary matrix model of section \ref{sec:CSesLSZ}.\par
To arrive at \eqref{eq:finalZSLSZ} we have first shifted both sets of variables and then reflected the first set. If we go back to equation \eqref{eq:superspectra} and define $\mu_{2;r}, \nu_{2;r}$ from $\xi_r, \tilde{\xi}_r$ with opposite sign, we get to an expression analogous to \eqref{eq:finalZSLSZ} but with both $(x_1, \dots, x_{N_1})$ and $(w_1, \dots , w_{N_2})$ integrated over the same domain. In each case, we have to first change variables and then insert the supersymmetric Schur factorization identity.\par
Comparing with \cite{Kimura:2015dfa}, we rewrite the expression \eqref{eq:finalZSLSZ} in the form 
\begin{equation}
\label{eq:sLSZcompact}
\boxed{
	\mathbf{Z}_{\mathrm{sLSZ}} \left( \mu, \nu \right)  = \mathbf{A}_{ N_1, N_2} \left( \mu, \nu \right) ~\mathbf{Z}_{\mathrm{ABJ}} ~ \left\langle W_{\mu} W_{\nu} \ \Id_{(- \infty, - \gamma_1]^{N_1}} (x)  \Id_{[\gamma_2, \infty)^{N_2}} (w)  \right\rangle_{\mathrm{ABJ}} ,
}\end{equation}
where $\langle \cdots \rangle_{\mathrm{ABJ}}$ is the average in the ABJ matrix model \eqref{eq:ABJM}, and $\Id_{(- \infty, - \gamma_1]^{N_1}} (x)$ (respectively $\Id_{[\gamma_2, \infty)^{N_2}} (w)$) is the $N_1$-dimensional ($N_2$-dimensional) indicator function. 
Besides, $W_{\mu}$ is the trace of the holonomy of a superconnection along an equatorial circle inside $\mathbb{S}^3$, in the representation $\mu$ of the supergroup $U(N_1 \vert N_2)$, which describes a Wilson loop in ABJ(M) theory \cite{Drukker:2009hy,Lee:2010hk}. Previous works on averages of supersymmetric Schur polynomials over the ABJ(M) ensemble include \cite{Matsumoto:2013nya,Matsuno:2016jjp,Furukawa:2017rzo}.\par
The probabilistic interpretation of the result \eqref{eq:sLSZcompact} in terms of large deviations away from the equilibrium in the ABJ(M) matrix model with supersymmetric Schur insertions, follows directly from the discussion in subsection \ref{sec:fluctuations}. In particular
\begin{equation*}
	\frac{ \mathbf{Z}_{\mathrm{sLSZ}} \left( \mu, \nu \right)  }{ \mathbf{Z}_{\mathrm{ABJ}} } = \mathbf{A}_{ N_1, N_2} \left( \mu, \nu \right) \  \left\langle W_{\mu} W_{\nu}  \right\rangle_{\mathrm{ABJ}} \ \mathrm{Prob}_{\mathrm{ABJ}; W_{\mu} W_{\nu}} ( x_{\mathrm{max}} \le - \gamma_1 , w_{\mathrm{min}} \ge \gamma_2  ) ,
\end{equation*}
hence the supermatrix model \eqref{eq:superLSZ} measures the probability of atypically large deviations from the equilibrium in the random matrix description of two supersymmetric Wilson loops carrying two $U(N_1 \vert N_2)$ typical representations. The specialization to two void partitions implies
\begin{equation*}
	\frac{ \mathbf{Z}_{\mathrm{sLSZ}} \left(\emptyset, \emptyset \right)  }{ \mathbf{Z}_{\mathrm{ABJ}} } =  \mathbf{A}_{ N_1, N_2} \left( \emptyset, \emptyset \right) ~  \mathrm{Prob}_{\mathrm{ABJ}} ( x_{\mathrm{max}} \le - \gamma_1 , w_{\mathrm{min}} \ge \gamma_2 ) .
\end{equation*}\par
\begin{remark}
Throughout this section we defined and analyzed a supermatrix version of the LSZ model. 
It would be interesting to derive the supermatrix model \eqref{eq:superLSZ} from an extension of the LSZ scalar theory to a noncommutative superspace.
\end{remark}

	\par
	\vspace{0.8cm}
	\subsection*{Acknowledgements}	
	
The work of MT was partially supported by the Funda\c{c}\~{a}o para a Ci\^{e}ncia e a Tecnologia (FCT) 
through its program Investigador FCT IF2014, under contract 
IF/01767/2014, during the first stages of this work. 
The work of LS was supported by the FCT through the doctoral scholarship 
SFRH/BD/129405/2017. 
The work is also supported by FCT Project 
PTDC/MAT-PUR/30234/2017. 
MT thanks Harold Steinacker and the Mathematical Physics group at University of Vienna, for discussions and warm hospitality during a short research visit at the end of 2016.

\appendix
\section{Schur polynomials}
\label{app:Schur}

Here we include explicit formulas involving Schur polynomials \cite{Macdonaldbook}.

\subsection{Spectral shift and rectangular Schur}

A simple identity of Schur polynomials quickly shows what occurs if, in the
case of a equispaced, harmonic oscillator spectra, we have a global overall
shift in the spectrum (that is, a different zero point energy). If we have a
rectangular partition of length $N$, $\left( l,l,...,l\right) $ which we
denote by $l^{N}$ then, assuming that $\lambda $ is a partition of
length equal or lower than $N$, it holds%
\begin{equation*}
s_{\mu +l^{N}}(e^{x_{1}},...,e^{x_{N}})=\prod_{i=1}^{N}e^{l x_{i}}s_{\mu }(e^{x_{1}},...,e^{x_{N}}),  \label{ext_rect}
\end{equation*}%
Therefore, an overall spectral shift by an integer $l$ in one external
matrix, corresponds to a renormalization of the mass parameter $\widehat{m}%
^{2} \mapsto \widehat{m}^{2}-l$.\par
Another useful identity is \cite{Macdonaldbook}
\begin{equation}
\label{eq:schurcid}
	s_{\mu} ( c e^{x_1}, \dots, c e^{x_N}) = c^{\vert \mu \vert } s_{\mu} ( e^{x_1}, \dots, e^{x_N}) .
\end{equation}

\subsection{Dimensions}

The value of $s_{\lambda }\left( 1,\dots ,1\right) $ gives the dimension of
the irreducible representation of $U(N)$ with highest weight $\lambda $.
Using Weyl's denominator formula 
\begin{equation*}
s_{\lambda }\left( 1,...,1\right) =\frac{\prod_{i<j}\left( \mu _{i}-\mu
_{j}\right) }{\prod_{i<j}\left( i-j\right) },
\end{equation*}%
where $\mu _{i}=\lambda _{i}+N-i$. Thus, it can also be written as 
\begin{equation*}
s_{\lambda }\left( 1,...,1\right) =\dim \lambda =\frac{1}{G(N+1)}%
\prod_{i<j}\left( \lambda _{i}-\lambda _{j}-i+j\right) ,
\end{equation*}%
where $G(\cdot)$ is the Barnes $G$-function.

\subsection{The Hopf link invariant}
\label{app:hopf}

We consider the Chern--Simons matrix model with two Schur insertions, and show that it computes the Hopf link invariant \cite{Kimura:2015dfa}. 
First, to deal with the inversion of variables appearing in a Schur polynomial, we can use the identity \cite{Macdonaldbook} 
\begin{equation*}
s_{\nu }\left( x_{1}^{-1},\dots ,x_{N}^{-1}\right)
=\prod_{j=1}^{N}x_{j}^{-\nu _{1}}s_{\nu ^{\ast }}\left( x_{1},\dots
,x_{N}\right) ,
\end{equation*}%
where the starred partition is defined as
\begin{equation*}
\nu ^{\ast }:=\left( \nu _{1}-\nu_{N},\nu _{1}-\nu _{N-1},\dots ,0\right) .
\end{equation*} 
Then the following holds: 
\begin{equation*}
\begin{aligned}
&\int_{[0, \infty)^N} \prod_{1 \le j<k \le N} \left( 2\sinh \left( \frac{z_{j}-z_{k}}{2}\right) \right)^{2} ~s_{\mu } \left( e^{z_{1}},\dots ,e^{z_{N}}\right) s_{\nu }\left( e^{z_{1}},\dots, e^{z_{N}}\right)  \prod_{j=1}^{N} e^{-\frac{1}{2g_{s}}z_{j}^{2}} \mathrm{d} z_{j}   \\
&=\int_{[0, \infty)^N} \prod_{1 \le j<k \le N} \left( 2\sinh \left( \frac{z_{j}-z_{k}}{2}\right) \right)^{2} ~s_{\mu }\left( e^{z_{1}},\dots ,e^{z_{N}}\right) s_{\nu ^{\ast}}\left( e^{-z_{1}},\dots ,e^{-z_{N}}\right) \prod_{j=1}^{N} e^{-\frac{1}{2g_{s}}z_{j}^{2} + \nu_1 z_j } \mathrm{d} z_{j}  \\
&=e^{\frac{N\nu _{1} ^2 g_{s}}{2}}\int_{[- \nu_1 g_s, \infty)^N} \prod_{1 \le j<k \le N} \left( 2\sinh \left( \frac{w_{j}-w_{k}}{2}\right) \right)^{2} ~ \prod_{j=1}^{N} e^{-\frac{1}{2g_{s}} w_{j}^{2}} \mathrm{d} w_{j}  \\
& \qquad\qquad \times s_{\mu }\left( e^{w_{1}+\nu _{1}g_{s}},\dots,e^{w_{N}+\nu _{1}g_{s}}\right) s_{\nu ^{\ast }}\left( e^{-w_{1}-\nu_{1}g_{s}},\dots ,e^{-w_{N}-\nu _{1}g_{s}}\right) \\
&=\mathcal{C}^{\prime \prime} _N \left( \vert \mu \vert, \vert \nu^{\ast} \vert \right) \int_{[- \nu_1 g_s, \infty)^N} \prod_{1 \le j<k \le N} \left( 2\sinh \left( \frac{w_{j}-w_{k}}{2}\right) \right)^{2} ~ \prod_{j=1}^{N} e^{-\frac{1}{2g_{s}} w_{j}^{2}} \mathrm{d} w_{j} \\
& \qquad \qquad \qquad \quad \times s_{\mu }\left( e^{w_{1}},\dots,e^{w_{N}}\right) s_{\nu ^{\ast }}\left( e^{-w_{1}},\dots ,e^{-w_{N}}\right) ,
\end{aligned}
\end{equation*}%
where $\mathcal{C}^{\prime \prime}_N \left( \vert \mu \vert, \vert \nu^{\ast} \vert \right) := e^{\frac{N\nu _{1} ^2 g_{s}}{2} + \nu _{1}g_{s} \left( \left\vert \mu \right\vert -\left\vert \nu ^{\ast}\right\vert \right) }$. 
Plugging this calculation in equation \eqref{eq:ZLSZgenY} provides an equivalent interpretation of $\mathcal{Z}_{\mathrm{LSZ}} (\mu, \nu)$ in terms of the Hopf link invariant carrying representations $\mu$ and $\nu^{\ast}$. We also note that, in this interpretation, setting $\nu=\mu$, which corresponds to two external matrices with equal eigenvalues, would not lead to the diagonal part of the Hopf link invariant, due to the appearance of the starred partition.\par
For the sLSZ model instead there is no equivalent interpretation in terms of the Hopf link with one starred partition. 
This is because, if we invert one set of variables in \eqref{eq:finalZSLSZ}, trading the partitions $\nu_1, \nu_2$ for $\nu_1 ^{\ast}, \nu_2 ^{\ast}$, the factorization of the supersymmetric Schur polynomial would bring one product with inverted variables, $\prod_{\ell, r} \left( e^{-x_\ell} + e^{-w_r} \right)$, and we would not arrive at the ABJ matrix model.\par
Recall that Chern--Simons observables depend on framing \cite{Witten:1989}:
the relation between such dependence and the modular matrices is given in 
\cite{Jeffrey:1992}. Observables in the matrix model description are not in
the canonical framing, and the (usual) Hopf link average is:%
\begin{equation*}
\left\langle W_{\mu \nu ^{\ast }}\right\rangle_{\mathrm{CS}} =\left( TST\right) _{\mu \nu^{\ast }}
\end{equation*}%
in terms of the modular $S,T$ matrices of the Wess--Zumino--Witten (WZW) model \cite{DiFrancesco:1997}. In the general case, the framed Hopf
link comes as $T^{n_1}ST^{n_2}$; if $n_1=n_2=0$, that is, the canonical framing in 
$\cS^{3}$, it is exactly the modular $S$ matrix, while if $n_1+n_2=2$, the
framing is the $U(1)$-invariant Seifert framing \cite{BlauThompson:2013}.
This is the case of the matrix model description.\par

	\bibliographystyle{ourstyle}
	\bibliography{NCCS_rev}
	
\end{document}